\documentclass[12pt,preprint]{aastex}

\shorttitle{DIG Halo Kinematics in NGC 4302}
\shortauthors{Heald et al.}

\newcommand{\kms}[1]{#1\,\mathrm{km\,s^{-1}}}
\newcommand{\kmskpc}[1]{#1\,\mathrm{km\,s^{-1}\,kpc^{-1}}}
\newcommand{\ha}{H$\alpha$}
\newcommand{\hi}{H$\,$I}
\newcommand{\nii}{[N$\,$II]}
\newcommand{\sii}{[S$\,$II]}

\newcommand{\zeone}{$-20\arcsec\,<\,z\,<\,-5\arcsec$}
\newcommand{\zzero}{$-5\arcsec\,<\,z\,<\,5\arcsec$}
\newcommand{\zwone}{$5\arcsec\,<\,z\,<\,20\arcsec$}
\newcommand{\zwtwo}{$20\arcsec\,<\,z\,<\,30\arcsec$}

\newcommand{\zeonepc}{$-1.6\,\mathrm{kpc}\,<\,z\,<\,-0.4\,\mathrm{kpc}$}
\newcommand{\zzeropc}{$-0.4\,\mathrm{kpc}\,<\,z\,<\,0.4\,\mathrm{kpc}$}
\newcommand{\zzeropcbm}{$z\,<\,0.4\,\mathrm{kpc}$}
\newcommand{\zwonepc}{$0.4\,\mathrm{kpc}\,<\,z\,<\,1.6\,\mathrm{kpc}$}
\newcommand{\zwtwopc}{$1.6\,\mathrm{kpc}\,<\,z\,<\,2.4\,\mathrm{kpc}$}
\newcommand{\npluss}{[N\,II]$\lambda\,6583$\,+\,[S\,II]$\lambda\,6716$}

\begin{document}

\title{Integral Field Unit Observations of NGC 4302:\\Kinematics of the Diffuse Ionized Gas Halo}
\author{George H. Heald\altaffilmark{1,2,3}, Richard J. Rand\altaffilmark{1,3}, Robert A. Benjamin\altaffilmark{4}, and Matthew A. Bershady\altaffilmark{5}}
\altaffiltext{1}{University of New Mexico, Department of Physics and Astronomy, 800 Yale Boulevard NE, Albuquerque, NM 87131}
\altaffiltext{2}{ASTRON, P.~O. Box 2, 7990 AA Dwingeloo, The Netherlands}
\altaffiltext{3}{Visiting Astronomer, Kitt Peak National Observatory, National Optical Astronomy Observatory, which is operated by the Association of Universities for Research in Astronomy, Inc. (AURA) under cooperative agreement with the National Science Foundation.}
\altaffiltext{4}{University of Wisconsin -- Whitewater, Department of Physics, 800 West Main Street, Whitewater, WI 53190}
\altaffiltext{5}{University of Wisconsin -- Madison, Department of Astronomy, 475 North Charter Street, Madison, WI 53706}

\begin{abstract}
We present moderate resolution spectroscopy of extraplanar diffuse ionized gas (EDIG) emission in the edge-on spiral galaxy NGC 4302. The spectra were obtained with the SparsePak integral field unit (IFU) at the WIYN Observatory. The wavelength coverage of the observations includes the \nii\,$\lambda\,6548,6583$, \ha, and \sii\,$\lambda\,6716,6731$ emission lines. The spatial coverage of the IFU includes the entirety of the EDIG emission noted in previous imaging studies of this galaxy. The spectra are used to construct position-velocity (PV) diagrams at several ranges of heights above the midplane. Azimuthal velocities are directly extracted from the PV diagrams using the envelope tracing method, and indicate an extremely steep dropoff in rotational velocity with increasing height, with magnitude $\kmskpc{\approx\,30}$. We find evidence for a radial variation in the velocity gradient on the receding side. We have also performed artificial observations of galaxy models in an attempt to match the PV diagrams. The results of a statistical analysis also favor a gradient of $\kmskpc{\approx\,30}$. We compare these results with an entirely ballistic model of disk-halo flow, and find a strong dichotomy between the observed kinematics and those predicted by the model. The disagreement is worse than we have found for other galaxies in previous studies.

The conclusions of this paper are compared to results from two other galaxies, NGC 5775 and NGC 891. We find that the vertical gradient in rotation speed, per unit EDIG scale height, for all three galaxies is consistent with a constant magnitude (within the errors) of approximately $15-25\,\mathrm{km\,s^{-1}\,scaleheight^{-1}}$, independent of radius. This relationship is also true within the galaxy NGC 4302. We also discuss how the gradient depends on the distribution and morphology of the EDIG and the star formation rates of the galaxies, and consequences for the origin of the gas.
\end{abstract}


\section{Introduction}\label{section:introduction2}

\setcounter{footnote}{2}

Gaseous halos are potentially excellent laboratories for the study of important aspects of spiral galaxy evolution. External, and some internal, processes have significant \emph{extra}planar manifestations. Interactions between the galaxy and its surrounding intracluster medium \citep[e.g.,][]{vcbd01}, intergalactic medium \citep[such as primordial accretion; e.g.,][]{o70}, and/or neighboring galaxies \citep[e.g.,][]{mlk98}, can obviously have effects outside of the disk. Internal processes, too, can have an impact on the environment beyond the star forming disk, such as galactic winds \citep[e.g.,][]{m03,vcb05}, and phenomena described by galactic fountain \citep{sf76,b80} or chimney \citep{ni89} models. An understanding of all these processes and their effects on galaxy evolution requires an understanding of the properties of the material surrounding the disk.

In the Milky Way, extraplanar gas has been discovered both in emission \citep[e.g.,][]{bkb85,wv97,kwmhb98,lp05,m05} and absorption \citep[e.g.,][]{fuse00,sswrmjsms03}. It is difficult to interpret these observations because our position within the Galaxy makes distance determinations difficult, and causes confusion along the line of sight; therefore, external systems are often targeted to enable study of the entire disk-halo interface. Gas has been found in the halos of many external spiral galaxies in the form of neutral hydrogen \citep[e.g.,][]{ssv97,mw03,foss05}, hot X-ray gas \citep[e.g.,][]{bp94,shchw04,tprbd06}, and diffuse ionized gas (DIG) \citep[e.g.,][]{rkh90,d90,r96,rd03a,mv03a}. Together with extraplanar dust \citep[e.g.,][]{hs99,im06} and extraplanar star formation \citep{trebfd03}, these observations paint a picture of a multiphase ISM extending well above the star forming disk.

The origin and evolution of this extraplanar ISM are not yet clear, but two alternatives are generally considered. First, the gas could have been accreted from the IGM \citep[e.g.,][]{o70}, or from companion galaxies \citep[e.g.,][]{vs05}. Alternatively, the gas could be participating in a star formation-driven disk-halo flow, such as the one described by the fountain model. Lines of evidence which link the morphological and energetic properties of the extraplanar DIG (EDIG) to the level of star formation activity in the underlying disk \citep[see, for example,][]{r96,hwr99,rd03a} would seem to support the latter idea, but clear examples of the former \citep[see the examples in][]{vs05} are observed. It is not yet clear how important the accretion process may be in systems with less obvious signs of external influence. To shed light on this question, additional evidence may be gained by studying the \emph{motion} of the halo gas.

In recent years, several groups have investigated the kinematics of extraplanar gas. The \hi\ halos of several galaxies have been found to rotate more slowly than the underlying disk: NGC 5775 \citep{lidcgw01}, NGC 2403 \citep{fosv01}, NGC 891 \citep{ssv97,foss05}, and possibly the low surface brightness galaxy UGC 7321 \citep{mw03}. Early studies of extraplanar rotation in the optical \citep[e.g.,][]{r00,tdsur00,mv03b} were limited by the long-slit spectra, in the sense that the one-dimensional observations did not allow for disentanglement of the density and velocity information encoded in the line profiles. Recent work \citep[][hereafter Papers I and II respectively]{hrbch06,hrbb06} has benefited greatly from the ready availability of integral field units (IFUs), which allow spectra to be obtained simultaneously over a two-dimensional portion of the sky. With these observations, the effects of density and velocity can be decoupled. In Paper I, TAURUS-II Fabry-Perot observations were used to establish that the EDIG in NGC 5775 has a gradient in rotational velocity with height above the midplane ($z$) of $\kmskpc{\approx 8}$; in Paper II, analysis of SparsePak observations of NGC 891 showed a gradient of $\kmskpc{\approx 15}$ in the NE quadrant of the halo. The latter result agrees with \hi\ results reported by \citet{foss05}.

To begin to assess whether the extraplanar gas has an internal origin, the kinematics of the EDIG in both NGC 5775 and NGC 891 have been compared to the results of a ballistic model \citep[developed by][]{cbr02} of disk-halo flow. Both galaxies are observed to have a steeper rotational velocity gradient than predicted by the ballistic model \citep[the discrepancy described by][for the neutral halo of NGC 891, is in the same sense]{fb06}. The difference between observations and model is greater for NGC 891, which has a \emph{less filamentary} EDIG morphology than NGC 5775. In this paper, we address the EDIG kinematics of a third edge-on spiral galaxy with well-studied extraplanar gas, NGC 4302. This galaxy was selected because it is extremely edge-on (see below), and was found by \citet{r96} to have the most prominent layer of EDIG within his sample of nine galaxies. Moreover, the morphology of the EDIG in NGC 4302, which is distinctly different than that in either NGC 5775 or NGC 891 (see discussion below and in \S\ \ref{section:discussion}), enables a comparison of EDIG kinematics in morphologically distinct gaseous halos.

NGC 4302 is classified as Sc in the Third Reference Catalogue of Bright Galaxies \citep[RC3;][]{ddcbpf91}, but this classification is somewhat uncertain because the galaxy is very nearly edge-on. A bar is likely to be present; see, for example, \citet{ldp00}. NGC 4302 is a member of the Virgo Cluster, and has a nearby companion, NGC 4298, with some signs of an interaction \citep[see, e.g.,][and discussion in \S\ \ref{section:kinematics2}]{kk04}. In this paper we assume a distance to NGC 4302 of $D=16.8\,\mathrm{Mpc}$, after \citet{t88}. To estimate the inclination angle, we closely examined the Two Micron All Sky Survey \citep[2MASS;][]{twomass06} $J$ and $H$ images of NGC 4302, both of which show a clearly defined dust lane. Based on the apparent offset of the dust lane from the major axis, we estimate an inclination angle $i\approx 89\degr$, and adopt that value for the remainder of the paper. Small deviations from this value will not significantly alter the results of our paper. The dust lane is very pronounced in the optical bands, and individual features are observed far from the midplane \citep[up to $\approx 1.5\,\mathrm{kpc}$; see][]{hs99}. The possible effects of dust extinction are discussed where appropriate throughout the paper. A summary of galaxy parameters for NGC 4302 is displayed in Table \ref{table:n4302params}.

The EDIG morphology of NGC 4302 is smoother than that of both NGC 5775 and NGC 891. It has been observed as a part of several EDIG imaging surveys \citep[e.g.,][]{pbs94,r96,rd00}. Taken together, the survey images demonstrate the existence of an extremely faint, smooth EDIG layer reaching up to $z\,\approx\,25\arcsec$ ($2\,\mathrm{kpc}$), and with an apparently sharp radial cutoff at $R'\,\approx\,50\arcsec$ ($4\,\mathrm{kpc}$).\footnote{To avoid confusion, we use $R$ to represent galactocentric radii, and $R'$ for major axis distance, throughout the paper. We express distances in linear units (kpc) in the former case, and offsets in angular units in the latter.} A single faint plume, extending to $z\,\approx\,-9\arcsec$ ($-0.73\,\mathrm{kpc}$; we take $z<0$ on the east side of the disk), near the nucleus of the galaxy was first reported by \citet{pbs94}, and later confirmed by \citet{rd00}. We are not aware of any other distinct EDIG features (in contrast to the extraplanar dust extinction, which has a quite intricate filamentary appearance). We note that an extended dust distribution can significantly alter the appearance of the \ha\ emission, as \citet{hs00} point out in their investigation of NGC 891. However, in the case of NGC 4302, we consider it unlikely that the dust extinction has significantly altered the appearance of the EDIG in such a way as to result in the observed smooth distribution.

This paper is arranged as follows. We describe the observations and the data reduction steps in \S \ref{section:observations2}. An examination of the halo kinematics is presented in \S \ref{section:kinematics2}, and the ballistic model is compared to the results in \S \ref{section:ballistic2}. In \S\ \ref{section:discussion}, we compare the EDIG halo kinematics of NGC 4302 with those of NGC 5775 and NGC 891, and discuss implications regarding the origin of the EDIG in these systems. We briefly conclude the paper in \S \ref{section:conclusions2}.

\section{Observations and Data Reduction}\label{section:observations2}

\begin{figure}
\epsscale{0.75}
\plotone{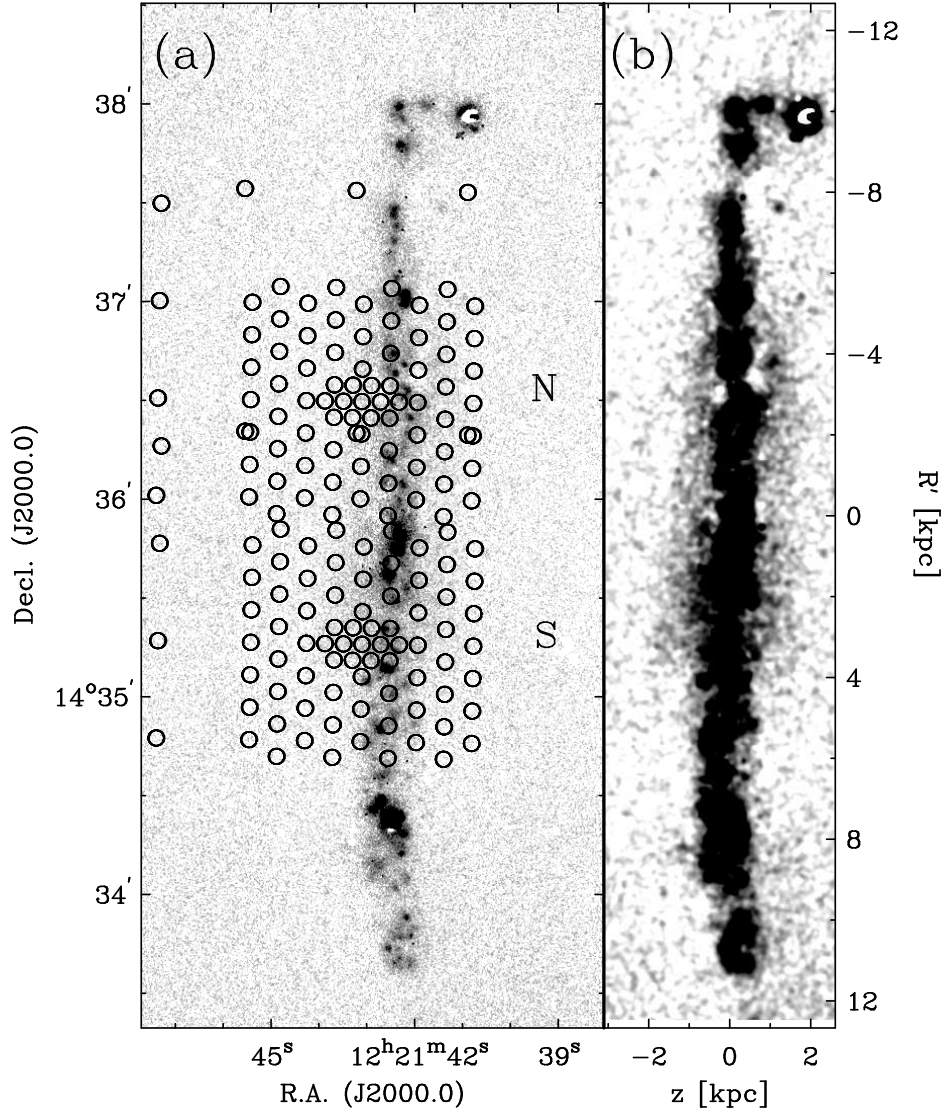}
\caption{(a) The two pointings of SparsePak, overlaid on the \ha\ image of NGC 4302 from \citet{r96}. The labels N and S indicate the IDs used for each pointing. The diameter of each SparsePak fiber is 4.687 arcsec. (b) The same image, smoothed to $1.5\arcsec$ in order to bring out the faint diffuse halo gas. The axes indicate the spatial scale at the adopted distance $D=16.8\,\mathrm{Mpc}$. Here and throughout the paper, $R'$ is positive on the receding (south) side, and $z$ is positive to the west.}
\label{fig:n4302sp}
\end{figure}

We used the SparsePak fiber array \citep{bahrv04,bavwcs05} to observe NGC 4302 during the nights of 2004 March 16-17 at the WIYN\footnote{The WIYN Observatory is a joint facility of the University of Wisconsin-Madison, Indiana University, Yale University, and the National Optical Astronomy Observatory.} 3.5-m telescope. SparsePak's 82 $4.687\arcsec$-diameter fibers fed the Bench Spectrograph, which was set up with the $860\,\mathrm{lines}\,\mathrm{mm}^{-1}$ grating at order 2. This setup yielded a dispersion of $0.462\,\mbox{\AA}\,\mathrm{pixel}^{-1}$ and a spectral resolution $\sigma_{\mathrm{inst}}=0.60\,\mbox{\AA}$ ($\kms{27}$ at \ha). The wavelength range covered the \nii\ $\lambda\lambda 6548,6583$, \ha, and \sii\ $\lambda\lambda 6716,6731$ emission lines. In Figure \ref{fig:n4302sp}, the two pointings are shown, overlaid on the \ha\ image of NGC 4302 from \citet{r96}. An observing log is presented in Table \ref{table:obslog2}.

The pointings were chosen to cover the brightest EDIG emission in the galaxy. The layer reported by \citet{r96}, defined by $|z|\,\leq\,25\arcsec$ ($2\,\mathrm{kpc}$) and $|R'|\,\leq\,50\arcsec$ ($4\,\mathrm{kpc}$) was within the coverage obtained with our two pointings (see Table \ref{table:obslog2}). The spectrograph setup described above allowed us to detect the EDIG, which is quite faint. As reported by \citet{r96}, at a height of $z=1\,\mathrm{kpc}$, the EDIG emission measure is $5-10\,\mathrm{pc\,cm}^{-6}$, about 3 times fainter than the EDIG in NGC 891 at the same height. In our spectra, we readily detect the \ha\ emission line at $z\,=\,1.3\,\mathrm{kpc}$ with a typical peak intensity of $1.76\,\times\,10^{-18}\,\mathrm{erg\,s^{-1}\,cm^{-2}\,arcsec^{-2}}\,=\,0.88\,\mathrm{pc\,cm}^{-6}$, assuming $T\,=\,10^4\,\mathrm{K}$. A higher resolution setup \citep[as was used by][to observe the brighter EDIG in NGC 891]{hrbb06} would have yielded significantly improved velocity resolution, but would also have greatly lengthened the exposure times necessary to obtain useful signal-to-noise ratios in the velocity profiles at the same heights that we consider here.

The initial data reduction steps were performed in the normal way using the IRAF\footnote{IRAF is distributed by the National Optical Astronomy Observatories, which are operated by the Association of Universities for Research in Astronomy, Inc., under cooperative agreement with the National Science Foundation.} tasks CCDPROC (for bias corrections) and DOHYDRA. This second task performs the flat field correction, and calibrates the wavelength scale using observations of a CuAr lamp, which were obtained approximately once per hour during the observing run. To estimate the accuracy of the wavelength solutions, we calculated the centroids of night-sky emission lines using the IRAF task SPLOT. The rms variation in the line centroids was $0.02\,\mathrm{\AA}$, corresponding to approximately $\kms{0.8}$ at \ha. DOHYDRA also does a rough fiber throughput correction using flat field observations, and finally extracts the spectra. At this point, the {\it relative} aperture throughput has been corrected, but the {\it absolute} correction remains unknown. Before determining the flux calibration using observations of a spectrophotometric standard star (Feige 67), light which fell outside the central fiber because of atmospheric blurring of the point spread function or imperfect centering of the star on the fiber must be accounted for. To correct for these effects, \citet{bavwcs05} have empirically determined the appropriate gain corrections (see their Appendix C4) based on measurements of the amount of starlight in the central fiber and the six surrounding fibers (which sample the wings of the standard star's point spread function). We have measured the amount of spillover for each of our spectrophotometric standard observations (obtained approximately once per hour throughout the run), applied the appropriate gain correction, and completed the flux calibration using the IRAF task CALIBRATE. Based on the uncertainties in the gain correction described above, we estimate the accuracy of the flux calibration to be at about the $0.7\%$ level; the precision is estimated to be at the $2\%$ level based on the variations between the individual observations of Feige 67.

Once the spectra were fully reduced, the continuum and skyline emission were subtracted in the following way. First, short segments of the spectra, containing object emission lines, were clipped out. These segments were chosen to be short enough that the continuum was well defined and could be removed with a linear fit. Next, sky lines were subtracted using our implementation of the ``iterative clipping'' method described by \citet[][their Appendix D]{bavwcs05}. This method accounts for the inherent field-dependence of optical aberrations in spectrographs. In each wavelength channel, a low-order function (for these observations, quadratic) was fit along the slit dimension. In order to obtain a good fit, more data points are required than are available by only considering fibers which are entirely free of object emission in all wavelength channels. Instead, all 82 fibers are considered together, and an iterative clipping scheme is implemented to obtain the correct profile shape along the slit dimension (i.e., the slit function for the optical aberrations). This technique is best suited for observations where the object emission is Doppler-shifted to widely varying wavelengths in the different fibers. In that way, sky lines (which vary little from fiber to fiber) may be readily distinguished from object emission and subtracted out. Because of the locations of our pointings (one on the receding side and the other on the approaching side), the object emission in our data is not distributed widely in wavelength space. Thus, application of this sky subtraction method leads to some oversubtraction of object emission, because the iterative clipping routine is unable to reject all of the object signal, leading to an overestimation of the normalization of the fit. In the region including \ha\ and \nii\ emission, the oversubtraction was minimal; in the region including the \sii\ lines, the oversubtraction was more severe. We have corrected for this effect by subsequently subtracting the average of the (sky-subtracted) spectra corresponding to fibers farthest from the midplane of NGC 4302 (fibers 6, 17, 19, 36, 38, 39, and 57), which did not receive any object flux. This has the effect of correcting the normalization of the fitted slit function, and thereby adding the oversubtracted galaxy emission line flux back into the spectra. The technique seems to work well for these observations. We tested for the possibility that line shapes were significantly affected by the skyline subtraction by comparing \ha, \nii, and \sii\ profile shapes in each fiber (because sky lines fall on different parts of the object emission line profiles, the line shapes should be affected in different ways for each emission line). This analysis did not reveal any systematic line shape errors. Following the sky subtraction, residual continuum emission was subtracted from the spectra.

In \S\ \ref{section:kinematics2}, we will analyze position-velocity (PV) diagrams constructed from the spectra. The PV diagrams were constructed by calculating the position, in the galaxy frame, of each fiber, and sorting the spectra into the appropriate ranges of $R'$ and $z$. Each PV diagram contains spectra from a different range of $z$. We choose the separation of pixels in the $R'$ axis to match the fiber diameter, thus maximizing the continuity of the diagrams. Note that when multiple spectra occupy the same location in a PV diagram, the average is used.

Because the EDIG in NGC 4302 is faint, with a small scaleheight \citep[$\approx 550-700\,\mathrm{pc}$;][]{cr01}, the number of independent PV diagrams that can be generated is small. We were able to detect EDIG in PV diagrams created for each of the following ranges of $z$: \zeone\ (\zeonepc); the major axis \zzero\ (\zzeropc); \zwone\ (\zwonepc); \zwtwo\ (\zwtwopc). We constructed PV diagrams separately for the \ha\ line and the (equally-weighted) sum of the \nii $\lambda 6583$ and \sii $\lambda 6716$ lines. The sum is used to increase signal-to-noise in the spectra. The total intensity of \npluss\ is approximately equal to that of the \ha\ line near the midplane, and increases to almost double the \ha\ intensity at $z\,\sim\,2\,\mathrm{kpc}$ as the \nii/\ha\ and \sii/\ha\ ratios both increase with $z$ \citep[see, e.g.,][]{mv03b}. Note that if there were any line shape errors induced by the sky subtraction algorithm, they should appear as systematic differences between the results from each type of PV diagram. No such differences are observed.

\section{Halo Kinematics}\label{section:kinematics2}

In this section, we analyze the PV diagrams constructed from the SparsePak spectra in an attempt to characterize the kinematics of the EDIG in NGC 4302. First, the envelope tracing method \citep{sa79,sr01} is used to directly extract rotation curves from the PV diagrams. We also present comparisons between the data PV diagrams and artificial model galaxy observations. The widths of the velocity profiles along the minor axis are examined, and, finally, we investigate radial variations in the observed halo kinematics.

\subsection{Envelope Tracing Method}\label{subsection:envelope2}

For an edge-on galaxy such as NGC 4302, care must be taken when deriving rotation speeds, for the reasons described by, e.g., \citet{kv04}. Briefly, a line of sight (LOS) through an edge-on disk crosses many orbits. Each orbit has a different velocity projection onto the LOS, and thus contributes differently to the total emission line profile. In the case of circular orbits, only one of those projections is maximal: the orbit that is intercepted by the LOS at the line of nodes (i.e., the line in the plane of the sky along which $R=R'$). Therefore, only that contribution to the line profile directly provides information about the rotation speed. The envelope tracing method uses PV diagrams to extract that information and build up rotation curves, under the assumption of circular orbits. To be more specific, the rotational (or azimuthal) velocity, $V_{\mathrm{\phi}}(R)$, is found by determining the observed radial velocity furthest from systemic in the velocity profile at each $R'$ (after corrections for the velocity dispersion of the material, and the velocity resolution of the instrument). We consider the results of the envelope tracing method to be valid only at radii where the rotation curve is approximately flat. At inner radii, where the derived rotation curve is still rising, a rising rotation curve cannot be distinguished from a changing radial density profile \citep[see also][]{foss05}.

The details of the envelope tracing algorithm are described in Paper I, and are not repeated here. We use the following values of the parameters in eqns. 1 and 2 of Paper I. The edge of the velocity profile is located using $\eta=0.2$ and $I_{\mathrm{lc}}=3\sigma$ (where $\sigma$ refers to the rms noise), which represent a scaling factor on the peak intensity of the line profile and a definition of the noise floor, respectively \citep[see also][]{sr01}. We also set $\sqrt{\sigma_{\mathrm{gas}}^2+\sigma_{\mathrm{instr}}^2}=\kms{40}$ (determined using the method described in \S\ \ref{subsection:modeling2}), where $\sigma_{\mathrm{gas}}$ is the velocity dispersion of the DIG. With the last value, we assume that $\sigma_{\mathrm{gas}}=\kms{30}$ (recall that $\sigma_{\mathrm{instr}}=\kms{27}$). If this assumption is incorrect, all of the derived rotation speeds will be incorrect by an additive constant, but the derived \emph{gradient} will be unchanged.

Because dust extinction in the disk may prevent the velocity profiles from including information from the line of nodes, we have also calculated a major axis rotation curve from \hi\ observations of NGC 4302 (Chung et al., in prep.). The \hi\ observations were obtained with the National Radio Astronomy Observatory (NRAO) Very Large Array (VLA)\footnote{The National Radio Astronomy Observatory is a facility of the National Science Foundation operated under cooperative agreement by Associated Universities, Inc.} in C configuration; the beam size is $17\arcsec\times\,16\arcsec$ ($1.4\,\mathrm{kpc}\times\,1.3\,\mathrm{kpc}$) at a position angle $-59\degr$, and the velocity resolution is $\sigma_{\mathrm{instr}}=\kms{10.4}$. A major axis cut was extracted from the data cube and kindly provided to us by J.~Kenney; we used the resulting PV diagram to obtain an \hi\ rotation curve using the envelope tracing method. An \hi\ velocity dispersion of $\kms{10}$ was assumed. Note that the \hi\ rotation curve was also used to verify the systemic velocity, under the assumption that the receding and approaching sides have the same rotation speed ($\kms{\approx\,175}$). The best value of the systemic velocity was confirmed to be $\kms{1150}$.

\begin{figure}
\epsscale{1}
\plotone{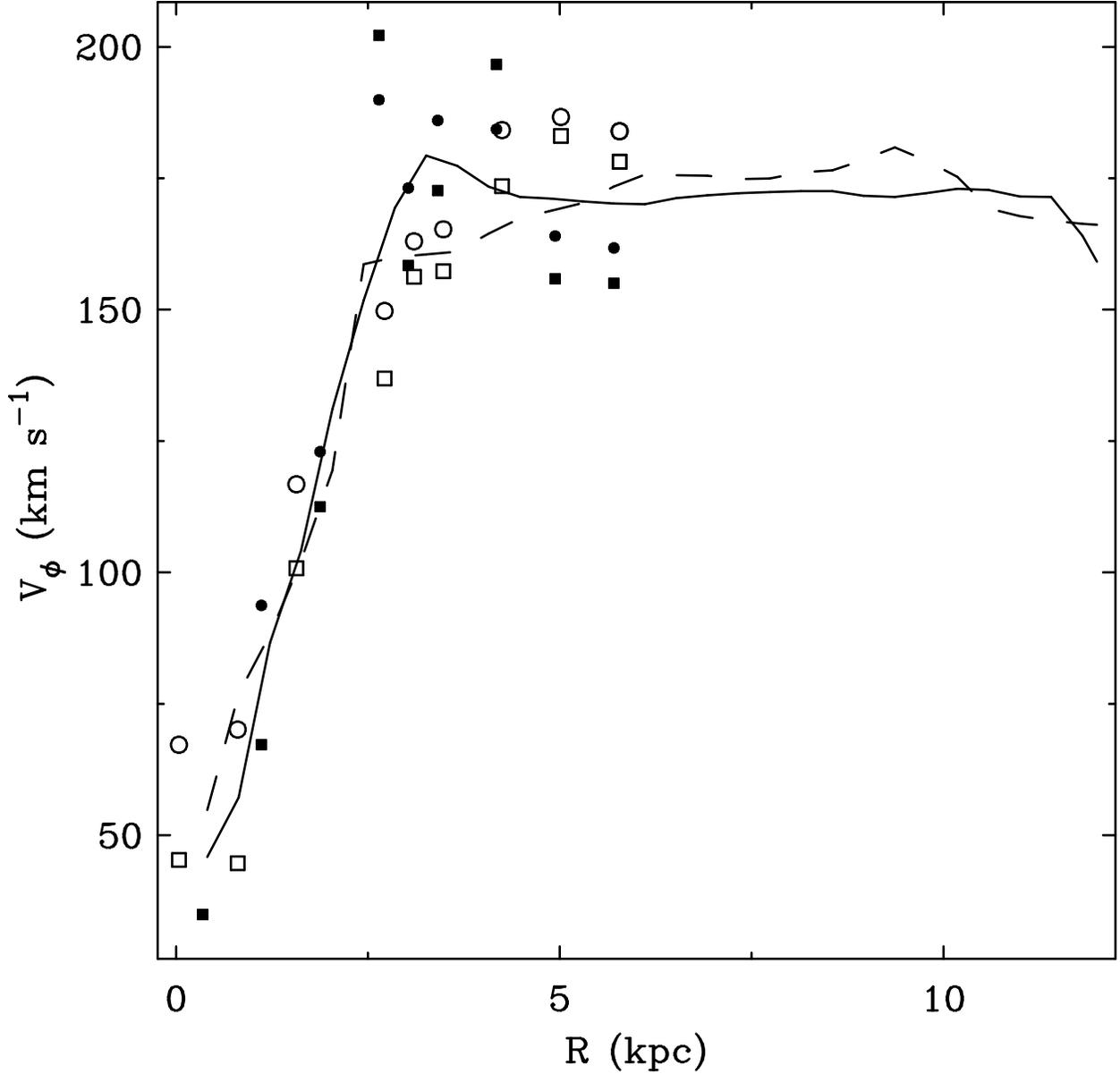}
\caption{Comparison of major axis rotation curves derived from the \hi\ (\emph{solid line}: approaching side; \emph{dashed line}: receding side), \ha\ (\emph{filled squares}: approaching side; \emph{open squares}: receding side), and \npluss\ (\emph{filled circles}: approaching side; \emph{open circles}: receding side) major axis PV diagrams. Typical errors are approximately $\kms{0.5-1}$ for the \hi\ rotation curves (for $R<10\,\mathrm{kpc}$), $\kms{0.5-2}$ for the \ha\ rotation curves, and $\kms{0.5-1.5}$ for the \npluss\ rotation curves.}
\label{fig:hidig}
\end{figure}

In Figure \ref{fig:hidig}, we compare the rotation curve derived from the \hi\ data to the one derived from the optical emission lines. Note that the radial extent of the \hi\ disk is much greater than the SparsePak coverage. The optical rotation curve shows a great deal more scatter, as can be expected since the \hi\ beam is much larger than the angular size of a SparsePak fiber. Still, the magnitude of the rotation curves appears to be approximately the same, and the asymmetry between the shapes of the \hi\ rotation curves on the approaching and receding sides seems to be reproduced in the optical. Extinction may explain some of the scatter in the optical rotation curve, and the apparently slower rotation speeds derived from the optical lines at $R\gtrsim\,5\,\mathrm{kpc}$ on the approaching (north) side of the disk. There are no other obvious signs that extinction may be systematically altering the optical rotation curves.

Having verified the value of the systemic velocity, and having found no evidence for significant extinction errors in the derivation of the rotation speeds, we next utilize PV diagrams constructed on the east and west sides of the disk to search for a change in average rotation speed with height above the midplane ($d\overline{V_{\mathrm{\phi}}}/dz$). In order to maintain consistency with previous papers, and to agree with Equation \ref{eqn:dvdzeqn4302} below, we adopt the convention that $d\overline{V_{\mathrm{\phi}}}/dz>0$ for rotation speed \emph{decreasing} with increasing height.

For the west side of the galaxy, azimuthal velocity curves were derived for PV diagrams constructed from spectra at \zwone\ and \zwtwo, and are shown in Figure \ref{fig:westrc}. A decrease in rotation speed with increasing height is apparent in the figure, though there is a large amount of scatter owing to the faint EDIG emission (the scale height, $7-9\arcsec$, is considerably less than the positions of the highest fibers, \zwtwo\ --- we are observing gas emission up to $\sim\,4$ scale heights from the disk), and perhaps some localized extinction. We nevertheless attempt to quantify the decrease in rotation speed with height, for the approximately flat part of the rotation curves. By determining the mean azimuthal velocity at each height for $R\gtrsim\,2\,\mathrm{kpc}$ (see Figure \ref{fig:westrc}), and fitting a linear relation to a plot of $\overline{V_{\mathrm{\phi}}}$-vs.-$z$, we have obtained values of $d\overline{V_{\mathrm{\phi}}}/dz$, which are listed in Table \ref{table:dvdztable}. We have also tabulated the gradients implied by the change in mean rotation speed from the midplane to the range \zwone\ (\zwonepc), and from there to \zwtwo\ (\zwtwopc). In order to ensure that the mean azimuthal velocities are not unfairly biased by values derived from spectra with low signal-to-noise ratios, we weighted the individual azimuthal velocities by the total signal-to-noise in the corresponding spectral line. There was no significant difference between the weighted and unweighted means; throughout this section, ``mean value'' refers to the weighted mean. Overall, the magnitude of the gradient seems to be approximately $\kmskpc{30}$.

We note that if extinction is preventing us from observing emission from the line of nodes, and therefore not truly measuring the azimuthal velocities, then our measured gradients will be \emph{lower} limits. The effects of extinction should diminish with increasing height, meaning that the line of sight probes farther into the disk as distance from the midplane increases. In a cylindrically rotating halo, such an effect would lead to an apparent \emph{rise} in rotation speed with height. Regardless, comparison with the \hi\ rotation curve indicated that extinction is not significantly affecting the derived rotation speeds.

On the east side of the disk, we are unable to trace the EDIG as far from the plane as on the west side. We are able to recover azimuthal velocities in the range \zeone\ (\zeonepc), but no higher. Because of the possibility that a gradient, if present, does not begin at $z=0$, we hesitate to draw conclusions from the absolute difference between azimuthal velocities in the midplane and in the range \zeone. Instead, we compare the azimuthal velocities to the east of the disk to those on the west side, to check whether the rotation appears to be consistent on both sides.

In Figure \ref{fig:ewcomp}, we compare the azimuthal velocity curves. Average azimuthal velocities for the approximately flat part of the rotation curves have also been calculated. The plot demonstrates very little difference between the rotation of EDIG on the east and west sides of the halo at the same absolute distance from the midplane, indicating that the west side (the side facing its companion, NGC 4298) does not have significantly different kinematics than the opposite side. This does not rule out a connection to a possible tidal interaction, but there is no evidence that the gaseous halo on one side of the disk is reacting differently than the other.

\subsection{PV Diagram Modeling}\label{subsection:modeling2}

The envelope tracing method is sensitive to the signal-to-noise ratio of the line profiles under consideration, and therefore the apparent gradient found above could be in part due to the falling signal-to-noise ratio with increasing height. In order to test for this possibility, and to ensure that the density profile of the gas is properly accounted for, we utilize another, independent method to derive the kinematics of the EDIG in NGC 4302. This method is based on constructing model galaxies with signal-to-noise ratios matched to the data at each $z$, performing artificial SparsePak observations of the models, and comparing modeled PV diagrams to the data.

To generate the galaxy models, we use a modified version of the Groningen Image Processing SYstem \citep[GIPSY;][]{vtbzr92} task GALMOD. GALMOD builds galaxies from a series of concentric rings. The properties of the rings are specified by: the radial density profile of emitting material; the major axis rotation curve; the velocity dispersion of the emitting material; the exponential scale height; and the viewing angle of the galaxy. Each parameter is allowed to vary from ring to ring, but with the exception of the radial density profile we force all parameters to be constant with radius. Note especially that by allowing the viewing angle to vary with radius, a warp can be built into the disk. The signature of a warp could be confused with a changing rotation speed with height \citep[see, e.g.,][]{ssv97}, but we assume that the star forming disk is not warped. Discretization noise is included in the model.

Our modification of the task allows for a linear variation in the rotation speed with increasing height, of the form
\begin{equation}
\label{eqn:dvdzeqn4302}
V_{\mathrm{\phi}}(R,z)=
\left\{
\begin{array}{ll}
V_{\mathrm{\phi}}(R,z=0) & \mbox{for $z\leq z_{\mathrm{cyl}}$} \\
V_{\mathrm{\phi}}(R,z=0)-\frac{dV}{dz}[z-z_{\mathrm{cyl}}] & \mbox{for $z>z_{\mathrm{cyl}}$}
\end{array}
\right.
\end{equation}
where $V_{\mathrm{\phi}}(R,z=0)$ is the major axis rotation curve, $dV/dz$ is the magnitude of the linear gradient (note that $dV/dz$ is positive for a declining rotation speed with increasing height), and $z_{\mathrm{cyl}}$ is the height at which the gradient begins. In general, we allow $z_{\mathrm{cyl}}$ to take nonzero values, but the present observations do not allow us to set any constraints on the value of this parameter, and we choose to leave it set to zero for the remainder of the modeling. The vertical (emission measure) scale height of the EDIG in NGC 4302 has been measured by \citet{cr01} to be $0.55-0.7\,\mathrm{kpc}$; for the models described here, the value $0.65\,\mathrm{kpc}$ was found to work well. The signal-to-noise ratio was matched to the data by adjusting the magnitude of the density distribution at the midplane.

The output of GALMOD is an artificial data cube. The noise is measured from the difference of two models constructed by varying only the random number seed. We have written a script which performs artificial SparsePak observations of the output data cubes, by extracting spectra in the pattern that the SparsePak fibers project onto the sky. Once the artificial observations have been made, the spectra are handled exactly like the data. The size of pixels along the velocity axis in the model is, by construction, the same as in the data, and the velocity resolution is set to be the same as in the data. In both the data and the model, PV diagrams are constructed separately for the approaching and receding sides of the disk and then joined together, except where noted.

To search for the best model for NGC 4302, we first attempted to estimate the radial density profile using the same technique employed in Papers I and II. In those investigations, the GIPSY task RADIAL was used to generate radial density profiles by considering intensity cuts in \ha\ images parallel to the major axis. In the case of NGC 4302, models created simply using pure exponential disks (scale length $7\,\mathrm{kpc}$, the same value used in \S\ \ref{section:ballistic2}), while unable to match some of the localized, small-scale features, were overall quite successful in matching the data. In retrospect, this result is expected because the EDIG distribution is extremely smooth and featureless \citep[see, e.g.,][]{rd00}. We therefore use the pure exponential radial density profile to generate our best-fit model. We also use a flat rotation curve ($\kms{175}$ based on the rotation curves derived from the \hi\ data; see Figure \ref{fig:hidig}), and a total velocity dispersion of $\kms{40}$ (as in the envelope tracing method).

Next, a sequence of models which vary only in their value of $dV/dz$ was generated. PV diagrams were constructed from the models as described above, for the same ranges of $z$ used to make the data PV diagrams. Difference PV diagrams are inspected, and the value of $dV/dz$ that minimizes the reduced chi-square ($\chi_{\nu}^2$), and the value that minimizes the mean difference between data and model, are chosen to define the best model for that particular data PV diagram. The results are presented in the first six rows of Table \ref{table:pvdvdz}. Although we consider the \ha\ and \npluss\ PV diagrams separately, in most cases the same model is compared with both. The exception is for \zwtwo, where the increased \nii/\ha\ and \sii/\ha\ line ratios require a higher amplitude (by a factor of 1.6) for the radial density profile in the \npluss\ model. Note that for the $z$-range \zwone\ (both in \ha\ and \npluss), clumpy structures in the data could not be fit well by an axisymmetric model. We therefore mask the three radii containing the largest deviations from axisymmetry in the PV diagrams.

The results of this analysis cluster around a gradient of approximately $\kms{30}$ kpc$^{-1}$, as did the envelope tracing analysis. There is a hint that the gradient is a bit higher than $\kmskpc{30}$. To demonstrate the quality of the PV diagram matching, we present overlays in Figures \ref{fig:biggridw} and \ref{fig:littlegride}, for the west and east sides, respectively. In the top two rows of Figure \ref{fig:biggridw} especially, it is difficult to judge by eye which model provides the best match to the data. Visual comparison is easiest for PV diagrams constructed from spectra at the highest $z$, where the leverage on $dV/dz$ is maximized. The variation in $\chi_{\nu}^2$ with the parameter $dV/dz$ is demonstrated in Figure \ref{fig:chisqzw2ns}, for the models listed in the first six rows of Table \ref{table:pvdvdz}.

Visual inspection of the overlay plots confirms that a vertical gradient in azimuthal velocity of magnitude $\kmskpc{30}$ provides the best match to the data, but in some cases the model with $dV/dz=\kmskpc{15}$ is also a good match. The conclusion from this analysis supports the result of the envelope tracing method, but there is some uncertainty in the exact value.

Because this technique relies on the model density profile being reasonably accurate to match the data, we test for the possibility that our adopted exponential disk model is biasing the results. To do this, we repeat the procedure described above, but using a flat radial density profile. This density profile does not match the data as well as the exponential disk, but we were able to analyze the statistics in the same way as before. The results are shown in the lower rows of Table \ref{table:pvdvdz}. Despite a large scatter due presumably to the poor match between data and model, and an apparent tendency for the best fit gradient to be somewhat lower than in the exponential disk models, these results are still consistent with $dV/dz=\kmskpc{30}$.

\subsection{Minor Axis Velocity Profile Widths}\label{subsection:minoraxis}

Our spectra can also be used to investigate other aspects of the gas kinematics in the halo of NGC 4302. Along the minor axis in a galaxy viewed from an edge-on perspective, the rotational velocity vectors will all be perpendicular to the line of sight (under the assumption of circular rotation). Hence, the only contributions by bulk motions to the width of velocity profiles on the minor axis will be made by radial motions. We have examined the spectra along (and close to) the minor axis in NGC 4302 to search for signs of radial motions. Using the GIPSY task XGAUPROF, Gaussian profiles were fitted to the individual velocity profiles (profiles which appeared non-Gaussian were neglected from consideration). The velocity dispersions resulting from this fitting procedure were corrected for the instrumental resolution ($\sigma_{\mathrm{V}}^2=\sigma_{\mathrm{fit}}^2-\sigma_{\mathrm{inst}}^2$) and are shown in Figure \ref{fig:sigmav}. The plotted errorbars are formal errors reported by the fitting routine. There appears to be no evidence for significant change in the widths of the velocity profiles as height above the midplane varies. Note that the apparent minimum in $\sigma_{\mathrm{V}}$ near $R'=0\arcsec$ is expected, because projections of the rotational velocity vectors onto the line of sight only vanish on the minor axis.

\subsection{Radial Variations in $dV/dz$}\label{subsection:dvdzr}

\begin{figure}
\epsscale{1}
\plotone{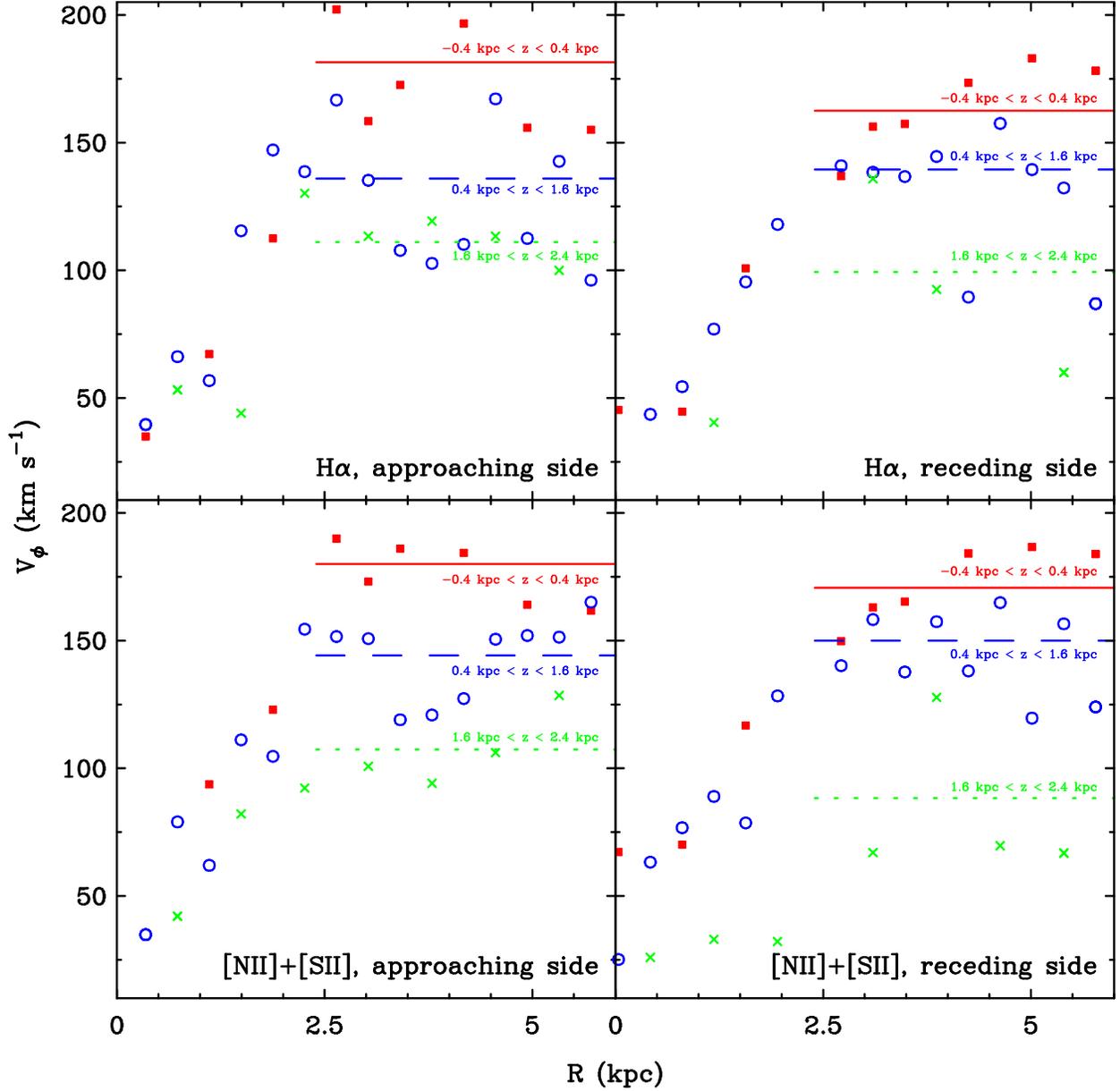}
\caption{Plots of azimuthal velocity curves, determined at each height, on the west side of the disk. The PV diagrams used to obtain these velocities were constructed from fibers at \zzeropc\ (\emph{filled squares} (colored red in electronic edition)), \zwonepc\ (\emph{open circles} (colored blue in electronic edition)), and \zwtwopc\ (\emph{crosses} (colored green in electronic edition)). The solid (colored red in electronic edition), dashed (colored blue in electronic edition), and dotted (colored green in electronic edition) lines respectively indicate the average azimuthal speed determined for each height, as described in the text. Azimuthal velocities were derived separately for the approaching side of the disk (\emph{left panels}) and the receding side of the disk (\emph{right panels}), using the \ha\ line alone (\emph{top panels}) and the sum \npluss\ (\emph{bottom panels}).}
\label{fig:westrc}
\end{figure}

\begin{figure}
\epsscale{1}
\plotone{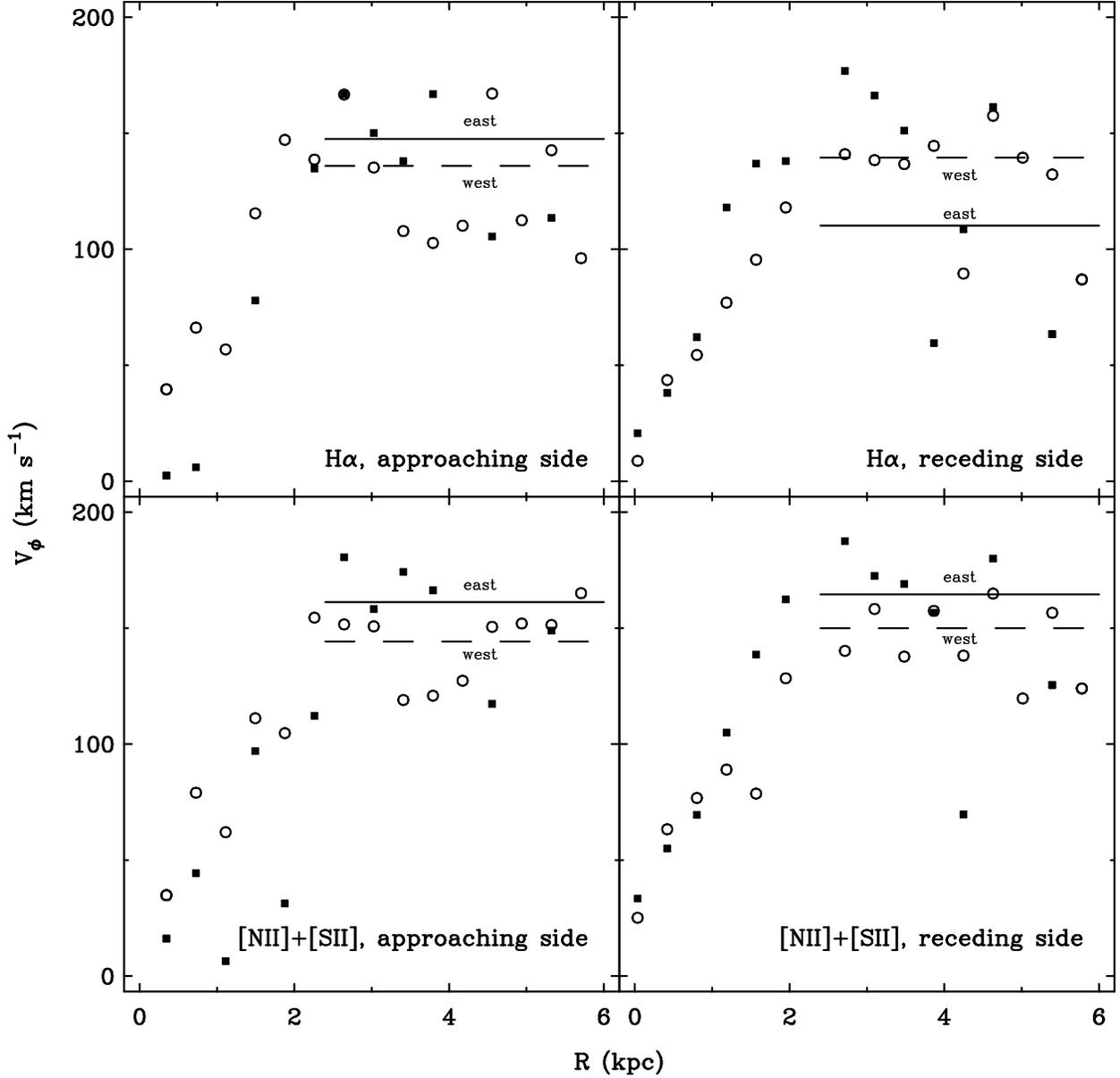}
\caption{Comparison of azimuthal velocities, as derived from PV diagrams constructed using spectra from the ranges \zeonepc\ (east; \emph{filled squares}), and \zwonepc\ (west; \emph{open circles}). Average azimuthal velocities for $R\gtrsim\,2.4\,\mathrm{kpc}$ are plotted as solid and dashed lines for the east and west $z$-ranges, respectively. Azimuthal velocities were derived separately for the approaching side of the disk (\emph{left panels}) and the receding side of the disk (\emph{right panels}), using the \ha\ line alone (\emph{top panels}) and the sum \npluss\ (\emph{bottom panels}).}
\label{fig:ewcomp}
\end{figure}

\begin{figure*}
\epsscale{1}
\plotone{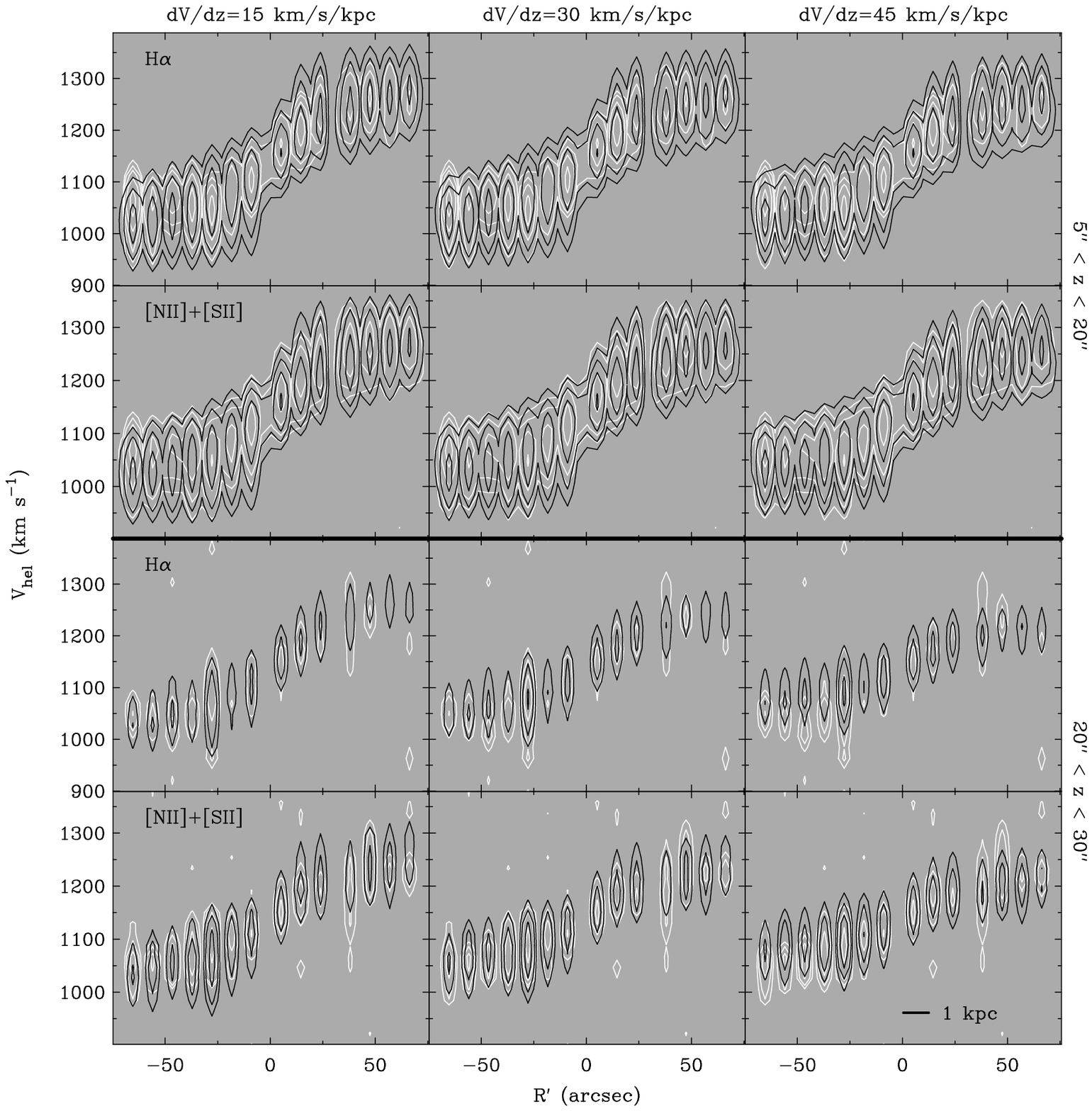}
\caption{Comparisons between PV diagrams constructed from the data (\emph{white contours}) and the galaxy models described in \S\ \ref{subsection:modeling2} (\emph{black contours}). The data PV diagrams were constructed from the \ha\ line (first and third rows) and the sum \npluss\ (second and fourth rows), on the west side, at the height ranges \zwone\ (top pair of rows) and \zwtwo\ (bottom pair of rows). The models were constructed using the base parameters described in the text, but different values of the gradient: $dV/dz=\kmskpc{15}$ (first column), $dV/dz=\kmskpc{30}$ (middle column), and $dV/dz=\kmskpc{45}$ (third column). Contour levels for both data and model are 8, 16, 32, 64, and $128\sigma$ in the first and second rows, and 2, 4, 8, and $16\sigma$ in the third and fourth rows, where $\sigma$ refers to the rms noise. Positive $R'$ is to the south. The systemic velocity is $\kms{1150}$. The spatial scale is indicated in the bottom right corner.}
\label{fig:biggridw}
\end{figure*}

\begin{figure}
\epsscale{0.5}
\plotone{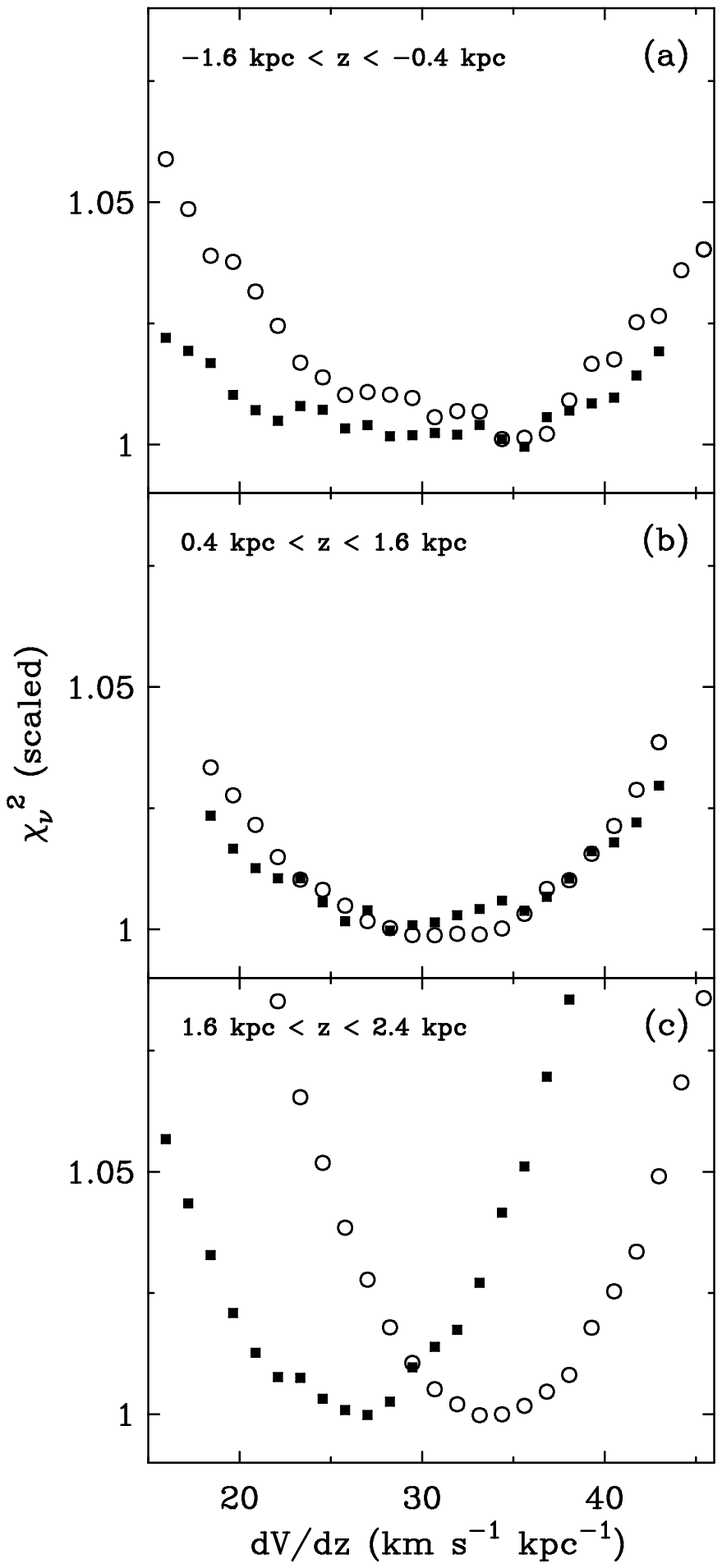}
\caption{Variation in the $\chi_{\nu}^2$ statistic with the parameter $dV/dz$, used to select the best (exponential-disk) model for PV diagrams constructed from the \ha\ line (\emph{filled squares}) and the sum \npluss\ (\emph{open circles}) at height ranges (a) \zeonepc, (b) \zwonepc, and (c) \zwtwopc. All $\chi_{\nu}^2$ curves have been divided by their minimum value for presentation; the minimum $\chi_{\nu}^2$ values are as follows: (a) 4.30 for \ha, 19.1 for \npluss; (b) 57.7 for \ha, 32.2 for \npluss; (c) 0.933 for \ha, 2.49 for \npluss. The minimum $\chi_{\nu}^2$ in each case occurred for the values of $dV/dz$ shown in Table \ref{table:pvdvdz}.}
\label{fig:chisqzw2ns}
\end{figure}

\begin{figure*}
\epsscale{1}
\plotone{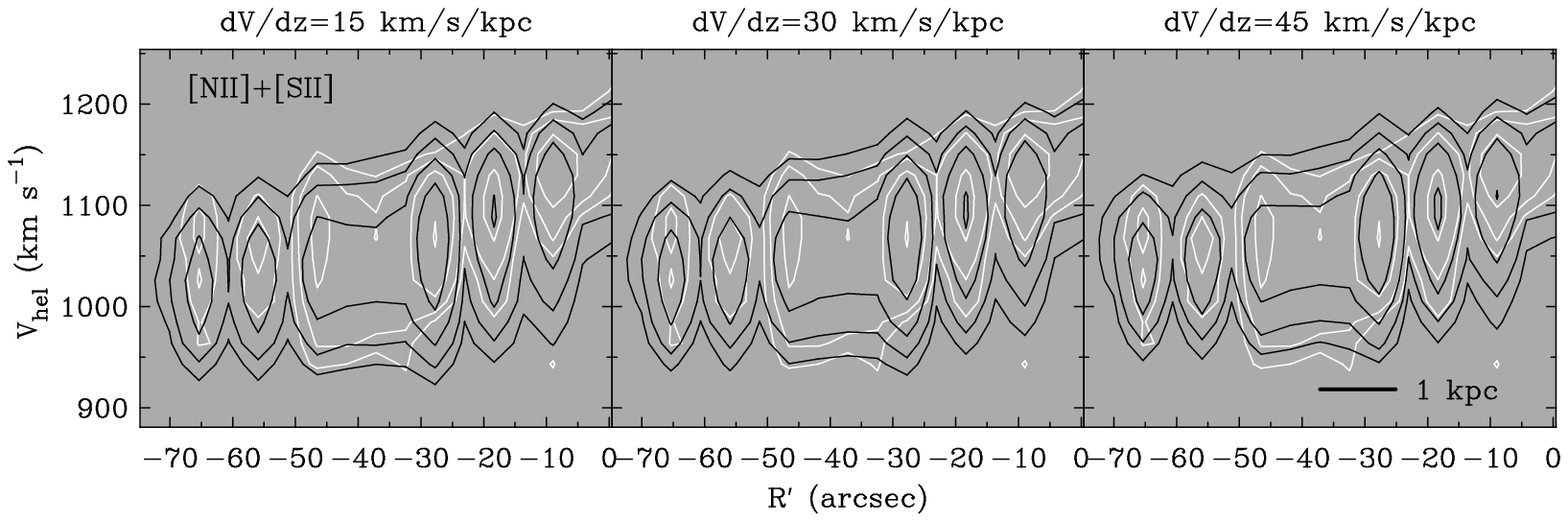}
\caption{Comparison between PV diagrams constructed from the data (\emph{white contours}) and the galaxy models described in \S\ \ref{subsection:modeling2} (\emph{black contours}). The data PV diagrams were constructed from the sum \npluss, at the height range \zeone\ (on the east side of the disk). The models were constructed using the base parameters described in the text, but different values of the gradient: $dV/dz=\kmskpc{15}$ (left), $dV/dz=\kmskpc{30}$ (middle), and $dV/dz=\kmskpc{45}$ (right). For the reasons described in the text, only data on the approaching side of the galaxy ($R'\,<\,0\arcsec$) are considered here. Contour levels for both data and model are 4, 8, 16, and $32\sigma$, where $\sigma$ refers to the rms noise. North is to the left. The systemic velocity is $\kms{1150}$. The spatial scale is indicated in the bottom right corner.}
\label{fig:littlegride}
\end{figure*}

\begin{figure}
\epsscale{0.6}
\plotone{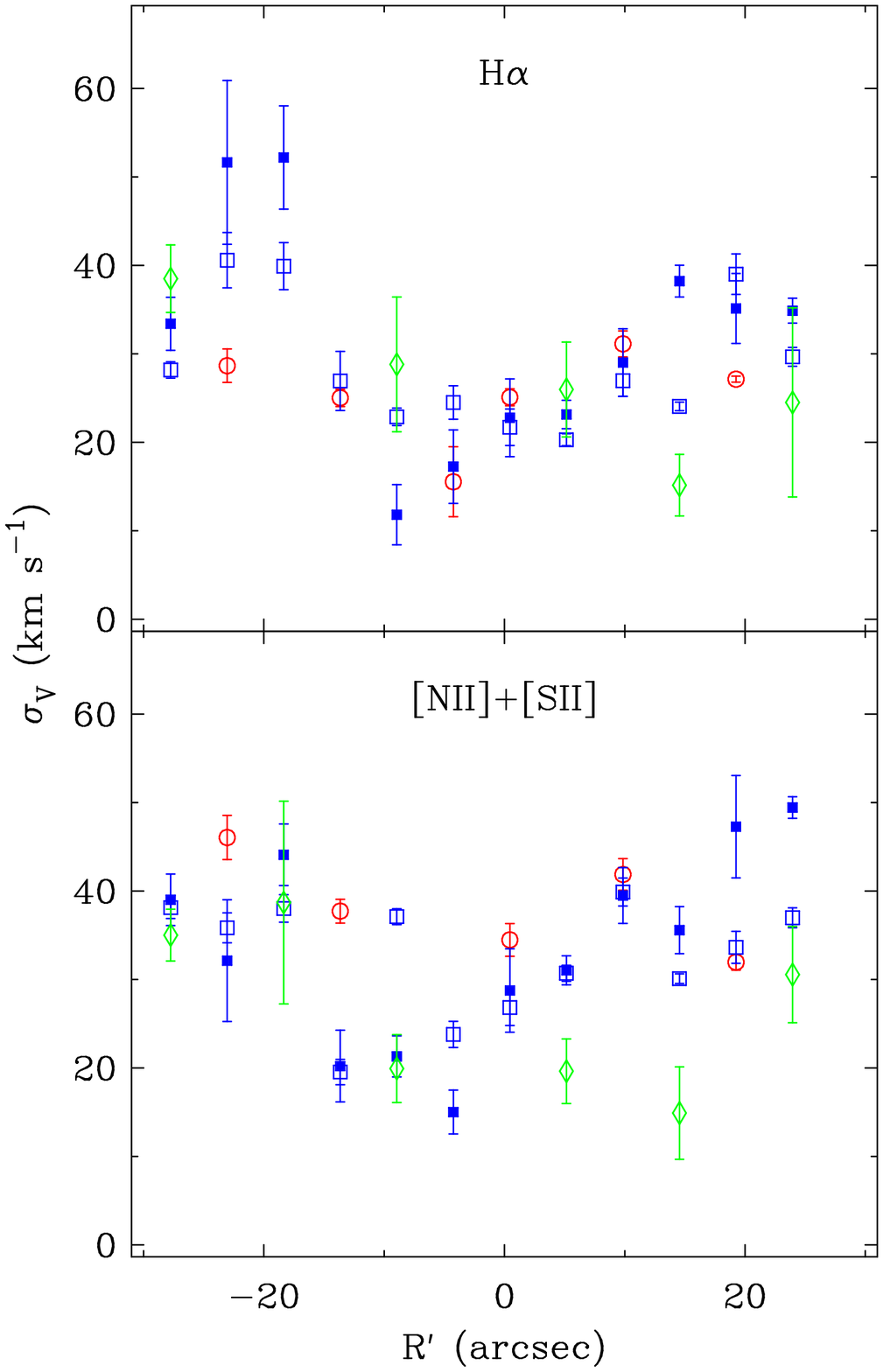}
\caption{Velocity dispersions, measured by fitting single Gaussian profiles to spectra along and near the minor axis of NGC 4302, and corrected for the instrumental resolution ($\sigma_{\mathrm{V}}^2=\sigma_{\mathrm{fit}}^2-\sigma_{\mathrm{instr}}^2$). Dispersions were measured for the \ha\ line (\emph{top panel}) and the sum \npluss\ (\emph{bottom panel}), at three different heights: \zzero\ (\emph{circles} (colored red in electronic edition)); \zeone\ (\emph{filled squares} (colored blue in electronic edition)); \zwone\ (\emph{open squares} (colored blue in electronic edition)); and \zwtwo\ (\emph{diamonds} (colored green in electronic edition)).}
\label{fig:sigmav}
\end{figure}

Until this point, we have treated the vertical gradient in rotational velocity as being independent of galactocentric radius. Yet an examination of Figure \ref{fig:westrc} suggests that at least on the receding side of the galaxy, the magnitude of the gradient may be variable, changing at a radius of about $4\,\mathrm{kpc}$. Moreover, as discussed by \citet[][see especially his Figure 13]{r96}, the \emph{appearance} of the EDIG layer changes at approximately the same radius; the EDIG is brighter at small radii ($R'\lesssim\,4\,\mathrm{kpc}$), and at outer radii the distribution seems to be more dominated by the H$\,$II regions. We come back to the relation between halo kinematics and morphology in \S\ \ref{section:discussion}; for now, we simply seek to justify treating the rotation curves inside and outside of $R\approx\,4\,\mathrm{kpc}$ separately.

To test for a different value of the gradient at different radii on the receding side of NGC 4302, we consider separately the rotational velocities displayed in Figure \ref{fig:westrc} in the radial ranges $2.5\,\mathrm{kpc}\,<\,R\,<\,4.25\,\mathrm{kpc}$ and $4.25\,\mathrm{kpc}\,<\,R\,<\,6\,\mathrm{kpc}$. (The radial break-point was chosen to be the midpoint of the full radial range; an additional test using a slightly different break-point, at $R\,=\,4.0\,\mathrm{kpc}$, yielded essentially the same results as are presented here.) Average rotation speeds at each height were calculated separately in both radial ranges. The runs of rotation speed versus height were well fitted with linear relationships, and the magnitudes of the fitted gradients are shown in Table \ref{table:dvdzr}. Clearly, there is a strong difference in the magnitude of the gradient when considering the two radial ranges separately. Inside $R=4.25\,\mathrm{kpc}$, the gradient is approximately a factor of 3 shallower than in the outer radial range.

We also performed the same experiment with the rotation speeds on the approaching side, which does not appear (from inspection of Figure \ref{fig:westrc}) to show a significant radial variation in velocity gradient. Indeed, the difference between the velocity gradients derived in each radial range using the \ha\ line is within the errorbars. In the case of the rotation speeds derived using the sum \npluss, there does appear to be a change in gradient from the inner disk to the outer disk, but in this case, it is likely to be due to extinction lowering the midplane rotation curve in the vicinity of $R'\approx\,5\,\mathrm{kpc}$ (this extinction was also invoked in \S\ \ref{subsection:envelope2} to explain the difference between the \hi\ and EDIG major axis rotation curves in Figure \ref{fig:hidig}). Additionally, separate measurements of the vertical scale height for the two radial halves of the approaching side are little different. We therefore conclude that there is no evidence for a radial variation in the velocity gradient or the vertical scale height on the approaching side, unlike the strong variation observed on the receding side.

\section{The Ballistic Model}\label{section:ballistic2}

The physical explanation for the vertical decrease in azimuthal velocity observed in NGC 4302, and other galaxies recently studied, is not yet understood. One possible scenario is described by the fountain model, which postulates that hot gas is lifted up into the halo by star formation activity in the disk. This hot gas would then cool and condense into clouds which move through the halo, and eventually rain back down onto the disk. As a first step toward understanding the dynamics implied by such a picture, a fully ballistic model of disk-halo cycling has been developed independently by two groups \citep{cbr02,fb06}. Models such as these, which consider the motion of non-interacting point masses in a gravitational potential, will be appropriate in cases where the density of the cycling clouds is sufficiently greater than that of the surrounding medium, so that their motion is essentially unperturbed by hydrodynamics. Whether the relative importance of hydrodynamics is indicated by an observable parameter, such as the morphology of the EDIG, is unknown, but it seems plausible that filamentary halo structures may imply a more ballistic disk-halo flow than smooth diffuse extraplanar gas layers. We return to this possibility in \S\ \ref{section:conclusions2}.

We have used the ballistic model of \citet{cbr02} in an attempt to model as closely as possible the kinematics of NGC 4302. The model numerically integrates the orbits of ballistic gas clouds in the galactic potential described by \citet{wmht95}. The clouds are initially placed in an exponential disk, and are launched vertically with an initial velocity randomly chosen to be between zero and a maximum ``kick velocity,'' $V_{k}$. As the clouds move upward and experience a weaker gravitational acceleration, they move radially outward. In order to conserve angular momentum, their azimuthal velocity drops as they move outward. Thus, a gradient in rotation speed is naturally included in the model. Together with the circular speed of the disk ($V_c$, which effectively sets the strength of the galactic potential), the kick velocity is the most critical parameter in setting the bulk kinematics of the clouds, as well as the scale height of the resulting steady-state halo of clouds. The clouds are assumed to have constant temperature, density, and size (and therefore equal \ha\ intensities); hence emission intensity is proportional to cloud column density along any line of sight. The interested reader should refer to \citet{cbr02} for a more complete description of the model.

The ballistic model we have considered is only one possible realization of the well known ``galactic fountain'' model. The dynamics of this type of circulation was first considered by \citet{b80} who calculated the trajectory of clouds that condense out of a hot, radiatively cooling halo. The difficulty in constructing such models lies in the lack of knowledge of the rotational structure of the gaseous halo that would form from flows of supernova heated gas. Bregman considered two cases, one in which gas flowing into the halo maintains the same specific angular momentum as it had in the disk (model ``a'' in his paper), and another in which pressure gradients in the halo lead to cylindrical rotation with the disk (model ``b''). Our ballistic model corresponds to Bregman's model ``a''. The cylindrical model ``b'' seems contraindicated by our observational results. More recent models of vertical circulation, e.g. \citet{ab04} and \citet{hth06}, have a more sophisticated treatment of the interstellar physics, but do not treat a large radial range, nor do they include rotation.

One of the ballistic model inputs is the scale length of the exponential disk. To estimate this, we first inspected a major axis intensity profile extracted from the \ha\ image of \citet{r96}. That profile suggested a scale length $R_{\mathrm{sc}}=7\,\mathrm{kpc}$, and the value was confirmed during the analysis described in \S\ \ref{subsection:modeling2} via comparisons between data and model PV diagrams. The circular speed is directly measured from the \hi\ rotation curve in Figure \ref{fig:hidig}: $V_c=\kms{175}$. The kick velocity is chosen so that the scale height of the model output is close to the measured scale height of the actual galaxy \citep[$550-700\,\mathrm{pc}$;][]{cr01}: $V_k=\kms{60}$. The model is run until the system reaches steady state (after $\approx\,1\,\mathrm{Gyr}$). The model outputs are the position and velocity of each cloud; these positions and velocities can then be used to generate an artificial data cube.

\begin{figure}
\epsscale{0.6}
\plotone{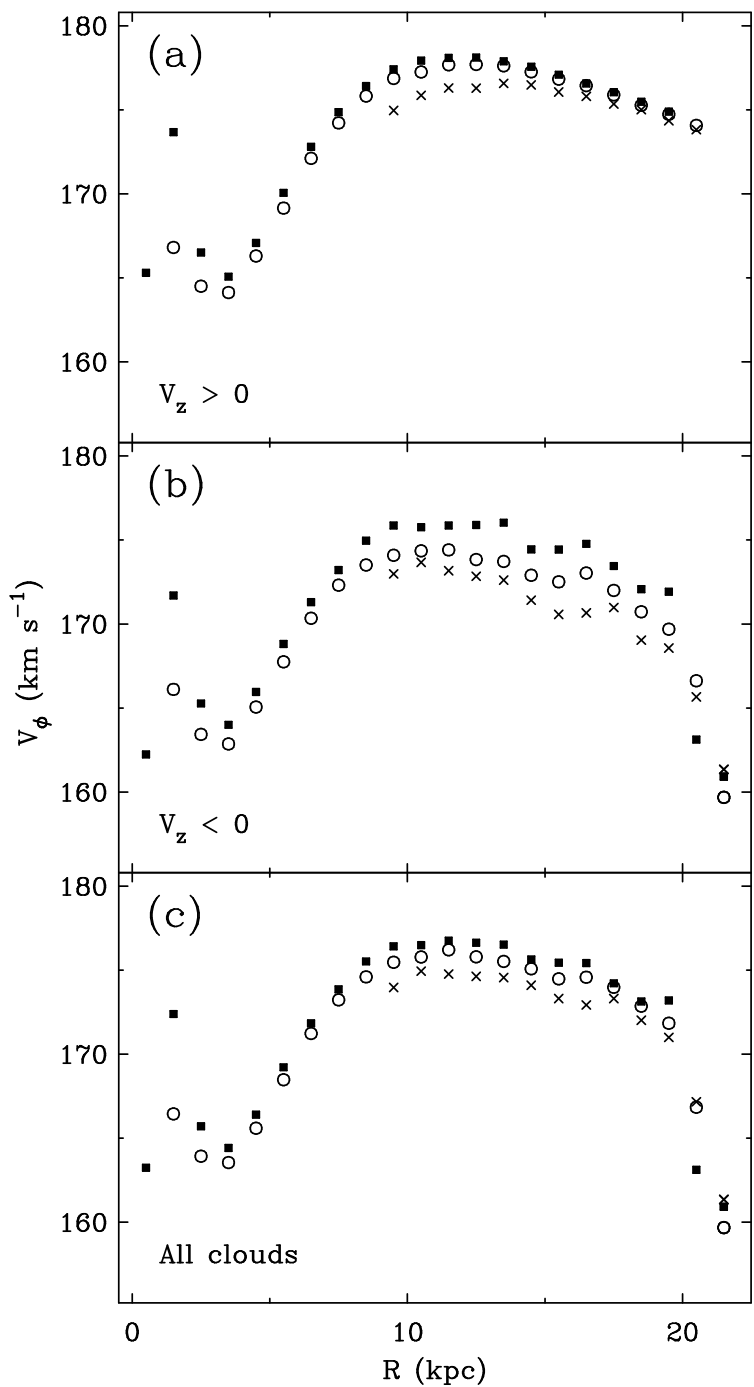}
\caption{Plots of azimuthal velocities extracted from the ballistic model. Rotation curves were created for clouds in the range \zzeropcbm\ (\emph{filled squares}), \zwonepc\ (\emph{open circles}), and \zwtwopc\ (\emph{crosses}). Separate rotation curves were extracted for (a) upward-moving clouds, (b) downward-moving clouds, and (c) all clouds.}
\label{fig:bmaz}
\end{figure}

The azimuthal velocities of the individual gas clouds can be directly extracted from the outputs of the ballistic model, and used to generate rotation curves. To compare the predictions of the ballistic model to the kinematics of the EDIG in NGC 4302, we have extracted rotation curves, using cloud velocities in the same height ranges considered in our analysis of the SparsePak data: \zzeropcbm, \zwonepc, and \zwtwopc. The results are shown in Figure \ref{fig:bmaz}. Note that we have extracted rotation curves for upward-moving and downward-moving (relative to the disk) clouds separately, in addition to considering all clouds together. The first condition would be appropriate if the clouds leave the disk as warm, ionized gas, but then cool and return as neutral gas. The second condition corresponds to a picture where the clouds leave the disk as hot gas, and return to the disk as warm ionized gas.

Inspection of Figure \ref{fig:bmaz} reveals immediately that the decrease in rotation velocity with height in the ballistic model is extremely shallow. The gradient, calculated in the radial range considered during the data analysis described in \S\ \ref{subsection:envelope2}, is approximately $\kmskpc{1.1}$ for upward-moving clouds, $\kmskpc{1.2}$ for downward-moving clouds, and $\kmskpc{1.0}$ for all clouds considered together. These gradients were calculated using only clouds in the vertical ranges \zzeropcbm\ and \zwonepc, because radial redistribution in the ballistic model leads to a lack of clouds in the range \zwtwopc\ for $R\lesssim\,10\,\mathrm{kpc}$, which is a larger radial range than is covered by the SparsePak data. We return to the issue of radial redistribution later.

The gradient in the ballistic model is far less than the value gleaned from the data, $\kmskpc{\approx\,30}$. Recall that the value of the gradient in the ballistic model is driven mainly by the value of the ratio of parameters $V_k/V_c$; in the case of NGC 4302, our best model (selected by matching the resulting exponential scale height to the data) is characterized by $V_k/V_c=60/175=0.34$. In order to attain a gradient with a higher magnitude, we might consider raising the value of the maximum kick velocity. However, a model with a kick velocity of $\kms{150}$ ($V_k/V_c=150/175=0.86$), while producing a gradient of up to $\kmskpc{\approx\,8}$, also generates a galaxy with a vertical scale height of $\approx\,31\,\mathrm{kpc}$. Clearly, increasing the kick velocity will never allow the ballistic model to match the vertical gradient in azimuthal velocity measured in the data, while simultaneously matching the scale height of the EDIG layer.

We also note that there is a great deal of flexibility in choosing the shape of the halo potential, without drastically affecting the kinematics of the clouds in this model. The reason for this, as illustrated in Paper II, is that the region in which the halo potential dominates the gravitational acceleration experienced by the clouds is outside of the area typically populated by the orbiting clouds. Hence, realistic changes to the halo potential will not dramatically change the orbital speeds of the ballistic particles.

\begin{figure}
\epsscale{0.6}
\plotone{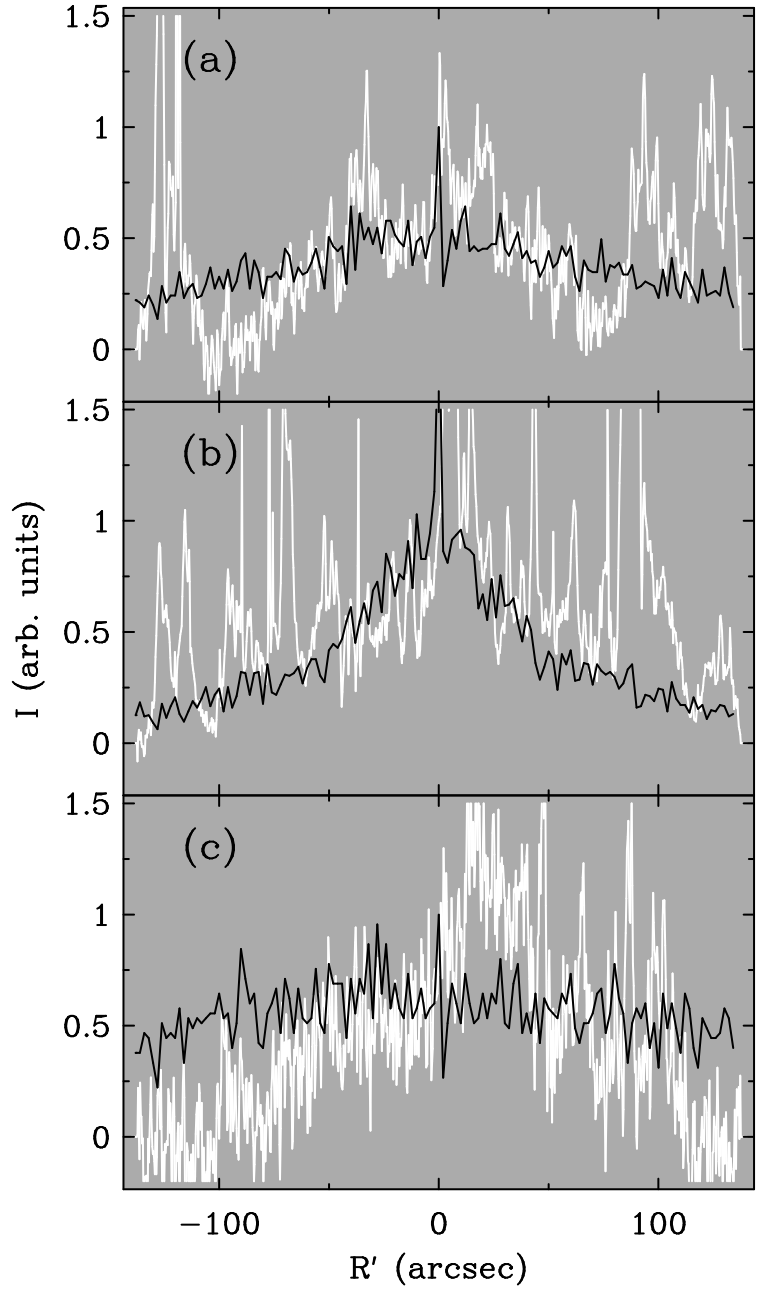}
\caption{Plots of intensity cuts parallel to the major axis in the the \ha\ image (\emph{white lines}) and the ballistic model (\emph{black lines}). Profiles are plotted for ranges (a) \zeone, (b) \zzero, and (c) \zwone.}
\label{fig:intcuts2}
\end{figure}

The plots in Figure \ref{fig:bmaz} are missing some data points in the range \zwtwopc\ because of a large amount of radial redistribution predicted by the ballistic model. But is the amount of radial redistribution predicted by the model borne out by the data? In Figure \ref{fig:intcuts2}, we present intensity cuts through the \ha\ image, as well as through a moment-0 (total intensity) map generated from the ballistic model output. The cuts are along the major axis and at two heights parallel to the major axis.

Comparison of the intensity cuts shows that while the distribution of clouds in the model at heights \zwonepc\ is relatively flat, the intensity distribution in the data is much steeper. In the \ha\ disk, the scale length of the intensity cut is approximately $7-8\,\mathrm{kpc}$; in the range \zwonepc, the scale length has dropped to about $2-5\,\mathrm{kpc}$, depending on location. Meanwhile, the scale length of the ballistic model intensity cuts are $\approx\,5\,\mathrm{kpc}$ in the disk, and about $10-20\,\mathrm{kpc}$ for \zwonepc. This shows that the outward radial migration in the ballistic model is excessive. The scale length of the radial intensity cuts could artificially appear to decrease with height in the data if extinction obscures emission from the central regions preferentially at lower $z$, but we consider this explanation unlikely.

\section{Discussion}\label{section:discussion}

Now that the ballistic model has been applied to a small sample of spirals with known, differing gradients in rotation speed with distance from the midplane, it may be appropriate to begin to look for patterns in order to guide future observations. We have summarized some important parameters of the galaxies in Table \ref{table:summary}. The listed EDIG scale heights ($h_z$) are electron scale heights, not emission measure scale heights (as are used in \S\ \ref{section:ballistic2}), and were determined by fitting (single-component) exponential profiles to \ha\ images of the three galaxies \citep[from][for NGC 5775, NGC 891, and NGC 4302, respectively]{crdw00,rkh90,r96}, over the ranges of $R$ and $z$ indicated in Table \ref{table:summary} (the ranges are comparable to those used to determine the values of $dV/dz$, which are also tabulated).

From the collection of data shown in Table \ref{table:summary}, some observed trends are intriguing. First, the tendency for the observed velocity gradient to diverge from the ballistic model predictions as the EDIG morphology becomes smoother and less filamentary may be an indication that smoother halos are intrinsically less governed by ballistic motion. Inversely, this lends support to the idea that halos with a more filamentary appearance are better described by models of ballistic motion. Effects which have been neglected thus far in our ballistic model, such as drag, may be important particularly in the smoother halos. However, we note that \citet{fbos07} have recently included drag in their ballistic model (the ballistic particles drag against a pre-existing, slowly rotating hot corona); although this addition provides a mechanism for removing angular momentum from the ballistic particles, thus better matching the \hi\ data for NGC 891, they find that the hot corona ``spins up'' after only $\sim\,1\,\mathrm{Myr}$, at which point the drag vanishes.

Another trend which may be suggestive is that the magnitude of the velocity gradient is seen to \emph{decrease} as the filamentary appearance of the halo, the EDIG scale height, and the level of star formation activity in the disk (as traced by the surface brightness of FIR luminosity, $L_{\mathrm{FIR}}/D_{25}^2$) each \emph{increase}. This would make sense in a picture where galaxies reside in halos populated by a variable mixture of accreting gas and star formation-driven fountain-type gas. In such a scenario, the gas associated with accretion, characterized on average by low angular momentum, should help to steepen the rotational velocity gradient \citep[e.g.,][]{fb06}; fountain-type gas, in the absence of such accreting gas, results in velocity gradients which are too shallow to reproduce the observed gas kinematics. Furthermore, this explanation of the trends we have found would also imply that halos with a more filamentary appearance contain proportionally more gas originating in the star forming disk, and less accreting low-angular momentum gas. Such halos are associated with disks with higher rates of star formation, which is traced here by the quantity $L_{\mathrm{FIR}}/D_{25}^2$. A larger sample of galaxies with varying EDIG morphologies, star formation rates, and observed rotation velocity gradients may be able to refine this picture, and help determine which parameters are most closely related.

Despite the tentative nature of these comparisons, the apparent connection between the magnitude of the velocity gradients, and the vertical EDIG scale heights, deserves additional comment. The product of the observed value of $dV/dz$ and the EDIG scale height, tabulated as $dV/dh_z$, has a large scatter ($\kms{14-36}\,\mathrm{scaleheight}^{-1}$), but in each case tends, within the errors, toward a value of $\kms{\approx\,20}\,\mathrm{scaleheight}^{-1}$. If confirmed in a larger sample of galaxies, this remarkable relationship should provide an excellent test for theoretical descriptions of the extraplanar gas (regardless of the details of the physical picture being modeled), as it would indicate a kinematic feature common to all gaseous halos. It is comforting that the range of EDIG scale heights in our small sample spans roughly a factor of 6, suggesting that we have covered a fairly large portion of the parameter space.

The correlations described above appear to not only hold true for whole galaxies as integrated units, but also \emph{within} individual galaxies. On the receding (south) side of NGC 4302, the vertical gradient in rotational velocity ($\kmskpc{31.0\,\pm\,19.8}$), expressed in units of the scale height ($0.56\,\pm\,0.01\,\mathrm{kpc}$), is $17.4\,\pm\,11.4\,\mathrm{km\,s^{-1}\,scaleheight^{-1}}$, consistent with the other measurements in Table \ref{table:summary}. But on that side of the disk, at a radius of about $4.25\,\mathrm{kpc}$, the characteristics of the EDIG change (this is clear both from inspection of the \ha\ image displayed in Figure \ref{fig:n4302sp}, and from consideration of the measured EDIG vertical scale heights). Beyond the same radius, the rotational velocity gradient of the halo gas becomes markedly steeper. Therefore, in Table \ref{table:summary}, measurements are tabulated separately for the two radial segments of the receding side of NGC 4302 (see also the discussion in \S\ \ref{subsection:dvdzr}). The variation happens in such a way as to keep the velocity gradient, per unit scaleheight, approximately constant.

One explanation for this growing body of observational data that seems firmly ruled out is the ballistic model. Other types of models \citep[e.g.,][]{bcfs06,kmwsm06} have explained previous analyses of NGC 891, but whether they are able to explain the near constancy of the velocity gradient per unit scaleheight remains to be seen. As discussed above, \citet{fb06} have speculated that the loss of angular momentum necessary to match the observations may be provided by an interaction between low-angular momentum, accreted gas and the pre-existing halo gas. Whether quantitative models of this process will reproduce the observed trends described here also remains to be seen. Clearly, more dynamical modeling of extragalactic infall, halo pressure gradients, and viscous interactions between the disk and halo would be valuable for interpretation of these results.

A major requirement of any successful model of NGC 4302 is that it not only reproduce the observed constancy of the velocity gradient per unit scaleheight between galaxies, but also the constancy of this quantity within a single galaxy with a radially varying scaleheight. This provides an intriguing challenge for future interpretation.

\section{Conclusions}\label{section:conclusions2}

We have presented SparsePak observations of the EDIG emission in the edge-on galaxy NGC 4302. By creating PV diagrams from the spectra, using both the \ha\ line and the sum \npluss\ independently, we extracted rotation curves at and above the midplane with the envelope tracing method, and used those azimuthal velocities to quantify the decrease in rotation speed with increasing height. The magnitude of the gradient appears to be approximately $\kmskpc{30}$, though there is a good deal of scatter (approximately $\kmskpc{10}$). Evidence for a steeper gradient beyond $R'\,\approx\,4.25\,\mathrm{kpc}$ is found on the receding side of NGC 4302.

As an alternative method for determining the variation in rotation speed with height above the disk, we have generated galaxy models constructed using different vertical gradients in azimuthal velocity, with signal-to-noise ratios matched to the data at each range of $z$ considered. Artificial observations of these models were used to construct PV diagrams, which were then compared to the data. Overall, the data tended to favor a gradient consistent with the envelope tracing results, $dV/dz\approx\,\kmskpc{30}$, but somewhat lower values are also possible.

The ballistic model of \citet{cbr02} was used to test the idea that the EDIG in NGC 4302 is taking part in a star formation-driven disk-halo flow. In the ballistic model, a radial outflow of clouds leads naturally to a decline in rotation speed with height. We extracted rotation curves directly from the output of the model, and comparison with the data revealed that the predicted gradient is far too shallow. It is possible to obtain higher values of the gradient by increasing the kick velocity, but this leads to an extremely large vertical scale height in the cloud distribution which is not observed in the data.

The ballistic model has an additional problem in reproducing the observations. Because the clouds undergo large-scale outward radial migrations, the distribution of clouds in strips parallel to the major axis rapidly flattens with increasing height, when viewed from an edge-on perspective. There is no evidence in the \ha\ image of \citet{r96} to suggest that this radial redistribution is taking place.

Finally, a comparison was made between the kinematics of the EDIG halos of NGC 4302, NGC 5775, and NGC 891. The steepest velocity gradient is measured in NGC 4302. It is intriguing that the kinematics of the halo gas appears to be related to other properties of the three galaxies: namely the morphology and vertical scale height of the EDIG halos, and the disk-averaged star formation rate (as traced by the quantity $L_{\mathrm{FIR}}/D_{25}^2$). The relationship between these parameters is in the sense that would be expected if the halos contain a variable mixture of star formation-driven fountain-type gas and low-angular momentum accreting gas, and that the steepness of the vertical gradient in rotational velocity is affected by the relative proportions of these sources of halo gas.

It is also very interesting to note that the present observations suggest that the decrease in rotational velocity \emph{per unit scale height} is approximately constant, with magnitude $\approx\,15-25\,\mathrm{km\,s^{-1}\,scaleheight^{-1}}$. This appears to hold true within the receding side of NGC 4302, where the velocity gradient and vertical scale height both change significantly, but in such a way that their product remains approximately constant.

\acknowledgments
We thank J.~Kenney for providing the \hi\ PV diagram, the observatory staff at NOAO/WIYN for their excellent support, and the anonymous referee for a comprehensive report which has improved the presentation of this paper. This material is based on work partially supported by the National Science Foundation under grant AST 99-86113.

\bibliography{ms}

\clearpage

\begin{deluxetable}{lcc}
\tabletypesize{\scriptsize}
\tablecaption{NGC 4302 Galaxy Parameters}
\tablehead{\colhead{Parameter} & \colhead{Value}}
\tablecolumns{2}
\startdata
R.A. (J2000.0)\tablenotemark{a} & 12 21 42.5 \\
Decl. (J2000.0)\tablenotemark{a} & +14 35 54 \\
Adopted distance\tablenotemark{b} & 16.8 Mpc \\
Adopted inclination & $89\degr$ \\
Position angle & $179\degr$ \\
Systemic velocity\tablenotemark{a,c} & $\kms{1150}$ \\
B-band luminosity ($L_B^0$)\tablenotemark{d} & $4.8\,\times\,10^{43}\,\mathrm{erg\,s}^{-1}$ \\
FIR luminosity ($L_{\mathrm{FIR}}$)\tablenotemark{e} & $<\,1.3\,\times\,10^{43}\,\mathrm{erg\,s}^{-1}$ \\
Star formation rate\tablenotemark{f} & $0.59\,M_{\sun}\,\mathrm{yr}^{-1}$ \\
Dynamical mass\tablenotemark{g} & $1.1\,\times\,10^{11}\,M_{\sun}$ \\
\enddata
\tablenotetext{a}{From NED.}
\tablenotetext{b}{From \citet{t88}.}
\tablenotetext{c}{Verified using \hi\ data, as discussed in \S\ \ref{subsection:envelope2}.}
\tablenotetext{d}{Based on the corrected apparent magnitude ($m_B^0$) given in \citet{ddcbpf91}.}
\tablenotetext{e}{NGC 4302 and its companion, NGC 4298, are not resolved in the IRAS survey, from which the FIR flux was taken.}
\tablenotetext{f}{Derived from $L_{\mathrm{FIR}}$ using the prescription given by \citet{k98}.}
\tablenotetext{g}{After \citet{hst82}, but adjusted to reflect the distance ($D=16.8\,\mathrm{Mpc}$) and inclination ($i=89\degr$) adopted in this paper.}
\label{table:n4302params}
\end{deluxetable}

\begin{deluxetable}{cccccccc}
\tabletypesize{\tiny}
\tablecaption{NGC 4302 SparsePak Observing Log}
\tablehead{\colhead{Pointing ID} & \colhead{RA\tablenotemark{a}} & \colhead{Decl.\tablenotemark{a}} & \colhead{Array PA} & \colhead{$R'$\tablenotemark{b}} & \colhead{$z'$\tablenotemark{b}} & \colhead{Exp. Time\tablenotemark{c}} & \colhead{rms Noise\tablenotemark{d}}\\ \colhead{(see Fig. 1)} & \colhead{(J2000.0)} & \colhead{(J2000.0)} & \colhead{($\degr$)} & \colhead{($\arcsec$)} & \colhead{($\arcsec$)} & \colhead{(hr)} & \colhead{(erg s$^{-1}$ cm$^{-2}$ \AA$^{-1}$)}}
\tablecolumns{8}
\startdata
N & 12 21 43.10 & 14 36 29.64 & $89$ & 
    $-72\,\mathrm{to}-3$ & $-43\,\mathrm{to}+24$ & 
    6.3 & $3.37\,(3.26)\,\times\,10^{-18}$\\
S & 12 21 43.10 & 14 35 15.96 & $89$ & 
    $+2\,\mathrm{to}+71$ & $-42\,\mathrm{to}+26$ & 
    6.8 & $3.22\,(3.05)\,\times\,10^{-18}$\\
\enddata
\tablenotetext{a}{R.A. and Decl. of fiber 52 (the central non-``sky'' fiber in the SparsePak array) for each pointing.}
\tablenotetext{b}{Ranges of $R'$ and $z'$ covered by each pointing of the fiber array. ``Sky'' fibers are not included in these ranges. $R'$ is positive on the south (receding) side, and $z'$ is positive to the west. At $D=16.8\,\mathrm{Mpc}$, $12\arcsec=1\,\mathrm{kpc}$.}
\tablenotetext{c}{Total exposure time, which is the sum of individual exposures of about 30 minutes each.}
\tablenotetext{d}{The rms noise was measured in the continuum near the \ha\ line for each of the 82 fibers in every pointing. The tabulated values are the mean (median).}
\label{table:obslog2}
\end{deluxetable}

\begin{deluxetable}{lccc}
\tabletypesize{\scriptsize}
\tablecaption{Summary of $dV/dz$ Values for the West Side, using Envelope Tracing Method}
\tablehead{\colhead{PV Diagram} & \colhead{$[d\overline{V_{\mathrm{\phi}}}/dz]_{\mathrm{fit}}$} & \colhead{$[d\overline{V_{\mathrm{\phi}}}/dz]_{0,1}$} & \colhead{$[d\overline{V_{\mathrm{\phi}}}/dz]_{1,2}$} \\
\colhead{(1)} & \colhead{(2)} & \colhead{(3)} & \colhead{(4)}}
\tablecolumns{4}
\startdata
App., \ha & $33\,\pm\,10$ & $45\,\pm\,27$ & $24\,\pm\,21$ \\
App., \npluss & $36\,\pm\,8$ & $35\,\pm\,16$ & $36\,\pm\,21$ \\
Rec., \ha & $29\,\pm\,14$ & $23\,\pm\,21$ & $39\,\pm\,28$ \\
Rec., \npluss & $33\,\pm\,14$ & $20\,\pm\,21$ & $61\,\pm\,31$ \\
\enddata
\tablecomments{(1) PV diagrams were constructed from fibers on the approaching (App.) or receding (Rec.) side, using either the \ha\ line or the sum \npluss.\\(2) Gradient determined from a linear fit to the average azimuthal velocity at all three heights (\zzeropc, \zwonepc, \zwtwopc), expressed in units $[\kmskpc{}]$.\\(3) Gradient determined from only the average azimuthal velocities at \zzeropc\ and \zwonepc, expressed in units $[\kmskpc{}]$.\\(4) Gradient determined from only the average azimuthal velocities at \zwonepc\ and \zwtwopc, expressed in units $[\kmskpc{}]$.}
\label{table:dvdztable}
\end{deluxetable}

\begin{deluxetable}{ccccc}
\tabletypesize{\scriptsize}
\tablecaption{Summary of Determinations of $dV/dz$ using PV Diagram Modeling Method}
\tablehead{\colhead{PV Diagram} & \colhead{Emission line(s)} & \colhead{Disk model} & \colhead{$[dV_{\mathrm{\phi}}/dz]_{\chi_{\nu}^2}$} & \colhead{$[dV_{\mathrm{\phi}}/dz]_{\mathrm{mean}}$} \\
\colhead{(1)} & \colhead{(2)} & \colhead{(3)} & \colhead{(4)} & \colhead{(5)}}
\tablecolumns{5}
\startdata
\zeonepc & \ha\tablenotemark{a} & Exp & $36$ & $31$ \\
\zeonepc & NS\tablenotemark{a} & Exp & $37$ & $36$ \\
\zwonepc & \ha & Exp & $28$ & $36$ \\
\zwonepc & NS & Exp & $31$ & $31$ \\
\zwtwopc & \ha & Exp & $27$ & $22$ \\
\zwtwopc & NS & Exp & $33$ & $34$ \\[3pt]
\zeonepc & \ha\tablenotemark{a} & Flat & $23$ & $34$ \\
\zeonepc & NS\tablenotemark{a} & Flat & $23$ & $15$ \\
\zwonepc & \ha & Flat & $22$ & $38$ \\
\zwonepc & NS & Flat & $18$ & $18$ \\
\zwtwopc & \ha & Flat & $12$ & $10$ \\
\zwtwopc & NS & Flat & $26$ & $26$ \\
\enddata
\tablecomments{(1) PV diagrams were constructed from fibers on both the approaching and receding side unless otherwise noted, in the listed ranges of $z$.\\(2) The data PV diagrams were constructed using either the \ha\ line or the sum \npluss\ (marked `NS' here).\\(3) The model PV diagrams were constructed using either an exponential or flat disk, as described in the text.\\(4) Gradient determined by minimizing the $\chi_{\nu}^2$ between data and model PV diagrams, expressed in units $[\kmskpc{}]$.\\(5) Gradient determined by minimizing the mean difference between data and model PV diagrams, expressed in units $[\kmskpc{}]$.}
\tablenotetext{a}{In the range \zeonepc, the spectra on the southern (receding) side were not able to be recreated with a realistic density profile, and are therefore not used for this analysis.}
\label{table:pvdvdz}
\end{deluxetable}

\begin{deluxetable}{ccc}
\tabletypesize{\tiny}
\tablecaption{Radial Variation in $dV/dz$}
\tablehead{\colhead{PV Diagram} & \colhead{Radial range} & \colhead{$[dV_{\phi}/dz]_{\mathrm{fit}}$}\\
\colhead{(1)} & \colhead{(2)} & \colhead{(3)}}
\tablecolumns{3}
\startdata
Rec., \ha & inner & $22.2\,\pm\,11.6$ \\
Rec., \npluss & inner & $23.3\,\pm\,12.9$ \\
Rec., \ha & outer & $59.5\,\pm\,5.0$ \\
Rec., \npluss & outer & $58.8\,\pm\,1.0$ \\[3pt]
App., \ha & inner & $33.4\,\pm\,7.2$ \\
App., \npluss & inner & $43.2\,\pm\,3.1$ \\
App., \ha & outer & $24.0\,\pm\,3.3$ \\
App., \npluss & outer & $14.7\,\pm\,3.4$ \\
\enddata
\tablecomments{(1) PV diagrams were constructed from fibers on the approaching (App.) or receding (Rec.) side, using either the \ha\ line or the sum \npluss.\\(2) The radial ranges $2.5\,\mathrm{kpc}\,<\,R\,<\,4.25\,\mathrm{kpc}$ (inner) and $4.25\,\mathrm{kpc}\,<\,R\,<\,6\,\mathrm{kpc}$ (outer) were considered separately.\\(3) Gradient determined from a linear fit to the average azimuthal velocity at all three heights (\zzeropc, \zwonepc, \zwtwopc), expressed in units $[\kmskpc{}]$.}
\label{table:dvdzr}
\end{deluxetable}

\setlength{\tabcolsep}{0.06in}
\begin{deluxetable}{c|c|c|c|c|c|c}
\tabletypesize{\tiny}
\tablecaption{Summary of Galaxy Parameters}
\tablehead{\colhead{Galaxy} & \multicolumn{2}{|c}{NGC 5775} & \multicolumn{1}{|c}{NGC 891} & \multicolumn{3}{|c}{NGC 4302}\\
\colhead{Halo Side/Quadrant} & \multicolumn{1}{|c}{SW side} & \multicolumn{1}{|c}{NE side} & \multicolumn{1}{|c}{NE quadrant} & \multicolumn{1}{|c}{NW quadrant} & \multicolumn{2}{|c}{SW quadrant}}
\tablecolumns{7}
\startdata
Observed $dV/dz\,[\kmskpc{}]$\tablenotemark{a} & $8\,\pm\,4$ & $8\,\pm\,4$ & $17.5\,\pm\,5.9$ & $34.5\,\pm\,6.4$ & $22.8\,\pm\,8.7$ & $59.2\,\pm\,2.5$ \\
$R$-range [kpc] & $0-12$ & $0-12$ & $4-7$ & $2.5-6$ & $2.5-4.25$ & $4.25-6$ \\
$z$-range [kpc] & $1.2-3.6$ & $1.2-3.6$ & $1.2-4.8$ & $0.4-2.4$ & $0.4-2.4$ & $0.4-2.4$ \\
\hline
$h_z\,[\mathrm{kpc}]$ & $2.2\,\pm\,0.07$ & $1.7\,\pm\,0.03$ & $1.40\,\pm\,0.02$ & $1.05\,\pm\,0.04$ & $0.63\,\pm\,0.02$ & $0.35\,\pm\,0.02$ \\
$R$-range [kpc] & $0-9.6$ & $0-9.6$ & $4-7$ & $2.5-6$ & $2.5-4.25$ & $4.25-6$ \\
$z$-range [kpc] & $2.4-4.3$ & $2.4-3.6$ & $0.5-2$ & $0.5-1.2$ & $0.5-1.1$ & $0.5-1.1$ \\
\hline
$dV/dh_z\,[\kms{}\,\mathrm{scaleheight^{-1}}]$ & $17.6\,\pm\,9.4$ & $13.6\,\pm\,7.0$ & $24.5\,\pm\,8.6$ & $36.2\,\pm\,8.1$ & $14.4\,\pm\,5.9$ & $20.7\,\pm\,2.1$ \\
\hline
Ball. mod. $dV/dz\,[\kmskpc{}]$\tablenotemark{b} & \multicolumn{2}{c|}{4} & $1-2$ & \multicolumn{3}{c}{1} \\
\hline
$L_{\mathrm{FIR}}/D_{25}^2\,[10^{40}\,\mathrm{erg\,s^{-1}\,kpc^{-2}}]$\tablenotemark{c} & \multicolumn{2}{c|}{8.1} & 2.2 & \multicolumn{3}{c}{$<\,2.3$\tablenotemark{d}} \\
\hline
EDIG Morphology\tablenotemark{c} & \multicolumn{2}{c|}{Many bright filaments} & Bright diffuse + filaments & \multicolumn{3}{c}{Faint diffuse} \\
\enddata
\tablenotetext{a}{From PV diagram modeling in the case of NGC 5775 (Paper I) and envelope tracing analysis in the cases of NGC 891 (Paper II) and NGC 4302 (this work).}
\tablenotetext{b}{Velocity gradients predicted by the ballistic model for NGC 5775 (Paper I), NGC 891 (Paper II), and NGC 4302 (this work).}
\tablenotetext{c}{From \citet{r96}.}
\tablenotetext{d}{NGC 4302 and its companion, NGC 4298, are not well resolved from each other in the IRAS survey, from which the values of $L_{\mathrm{FIR}}$ are derived.}
\label{table:summary}
\end{deluxetable}

\clearpage


\end{document}